\documentclass[a4paper,11pt]{article}
\pdfoutput=1 

\usepackage{jheppub} 

\usepackage[T1]{fontenc} 
\usepackage[english]{babel}
\usepackage{graphicx}
\usepackage{amsmath}
\usepackage{color}
\usepackage{xpatch}
\usepackage{caption}
\usepackage{subcaption}
\usepackage{slashed}
\usepackage{tikz}
\usepackage{multicol}
\usepackage{multirow}
\usepackage{comment}
\usepackage{hyperref}
\usepackage[capitalise]{cleveref}
\usepackage{makecell}
\usetikzlibrary{decorations.pathreplacing}
\usepackage[normalem]{ulem}

\newcommand{\eps}{\ensuremath{\epsilon}}

\newcommand{\as}{{\alpha_s}}

\def\cJ{\mathcal{J}}

\def\cS{\mathcal{S}}

\def\cM{\mathcal{M}}
\def\cA{\mathcal{A}}
\def\cF{\mathcal{F}}
\def\cO{\mathcal{O}}
\def\cR{\mathcal{R}}
\newcommand{\Tad}{\mathrm{Tad}}
\newcommand{\Bub}{\mathrm{Bub}}
\newcommand{\Tri}{\mathrm{Tri}}
\newcommand{\Boxm}{\mathrm{Box}}
\newcommand{\sfn}[2]{\mathrm{f}_{#1,#2}}
\def\sfvec{\vec{\mathrm{f}}}
\def\ZZ{\mathbf{Z}}
\def\dZ{\delta Z}

\def\la{\langle}
\def\ra{\rangle}
\def\spA#1#2{\la#1#2\ra}
\def\spB#1#2{[#1#2]}

\def\GeV{\mathrm{GeV}}
\def\TeV{\mathrm{TeV}}

\def\qb{\bar{q}}

\def\qqaa{0\to \bar{q}q\gamma\gamma}
\def\ggaa{0\to gg\gamma\gamma}
\def\qq{\mathrm{\bar{q}q}}
\def\gg{\mathrm{gg}}
\def\helccs{\tilde{A}}
\def\helccfin{\tilde{F}}
\def\helccpole{\tilde{P}}
\def\nplone{\mathrm{NPL1}}

\def\usepix#1#2#3#4#5#6{\parbox{#1}{\includegraphics[width=#1,trim= #3 #4 #5 #6]{#2}}}

\title{\boldmath 
Two-loop QCD amplitudes for $pp\to\gamma\gamma$ with a top-quark loop
}
\author[a]{Souvik Bera,}
\author[a,b]{Heribertus Bayu Hartanto}

\affiliation[a]{Asia Pacific Center for Theoretical Physics, Pohang, 37673, Korea}
\affiliation[b]{Department of Physics, Pohang University of Science and Technology, Pohang, 37673, Korea}

\emailAdd{souvik.bera@apctp.org}
\emailAdd{bayu.hartanto@apctp.org}

\abstract{
We revisit the computation of two-loop QCD amplitudes for diphoton production in the quark- and gluon-initiated channels, focusing on contributions involving a closed top-quark loop.
We derive analytic expressions for the finite remainders in terms of a set of special functions 
constructed from the $\epsilon$-expanded components of master integrals.
We employ a canonical basis of master integrals for the polylogarithmic sectors, while retaining a noncanonical basis for the elliptic sectors.
The analytic cancellation of ultraviolet and infrared poles is established by deriving relations among master integral components. 
Relations among polylogarithmic components are derived numerically from their differential equations and we utilise 
a set of finite Feynman integrals, constructed using local numerator insertions and constraints from soft and collinear limits, 
to derive relations between pole-contributing elliptic master integral components and the polylogarithmic ones.
The construction does not rely on a particular choice of noncanonical elliptic master integral basis.
The resulting set of special functions is minimal in the polylogarithmic sectors, whereas the elliptic 
components contributing to the finite part of the amplitudes are potentially overcomplete.
The special functions are evaluated numerically by solving systems of differential equations, enabling their computation across the physical phase space.
We present a \textsc{C++} implementation of the corresponding hard functions, suitable for on-the-fly numerical evaluation in cross-section calculations.
}

\begin{document} 
\maketitle
\flushbottom

\section{Introduction}
\label{sec:intro}

Collider events involving final-state photons offer clean signatures that can facilitate precision tests of the Standard Model of
Particle Physics (SM) and new physics searches. Production of two final-state photons, for instance, has been one of the central processes
studied at the Large Hadron Collider (LHC), particularly in the study of the Higgs boson. Precise measurements of
diphoton spectra have been carried out by LHC experiments~\cite{ATLAS:2012fgo,  
      ATLAS:2017cvh,
      ATLAS:2021mbt, 
      CMS:2014mvm, 
      ATLAS:2022fnp, 
      ATLAS:2023tnc, 
      CMS:2022wpo,
      CMS:2025fwn,
      CMS:2026weq} enabling stringent tests of perturbative QCD predictions across a wide kinematic range, supporting Higgs studies
and providing sensitive probes of resonant and non-resonant new-physics contributions at high invariant mass ($m_{\gamma\gamma}$).

Theoretical predictions for diphoton production at the LHC have been computed recently through next-to-next-to-next-to-leading order (N3LO) accuracy in QCD~\cite{Czakon:2026zhi}
improving upon the next-to-next-to-leading order (NNLO) QCD predictions which have been available for more than a 
decade~\cite{Catani:2011qz,Campbell:2016yrh,Catani:2018krb}.
While those high-precision calculations were done in massless QCD, top-quark mass ($m_t$) effects might be sizeable both in the high $m_{\gamma\gamma}$
and in the threshold region, $m_{\gamma\gamma} \simeq 2 m_t$~\cite{Maltoni:2018zvp,Chen:2019fla}. 
The complete NNLO QCD prediction for diphoton production including full top-quark mass dependence
was presented in Ref.~\cite{Becchetti:2023yat}. Such a prediction requires the computation of two-loop scattering amplitudes involving a closed top-quark loop in the quark-initiated
channel ($q\bar{q}\to\gamma\gamma$). 
The gluon-initiated counterpart formally contributes to the N3LO corrections since the corresponding leading-order amplitude is loop induced.
This class of massive scattering amplitudes is of particular interest due to the presence of elliptic Feynman integrals.
Unlike Feynman integrals that evaluate to multiple polylogarithms, elliptic Feynman integrals satisfy 
differential equations that cannot be brought into canonical form~\cite{Henn:2013pwa} by algebraic transformations, such that both finding the canonical basis and 
efficient numerical evaluation are in general more challenging.
Recent developments in the understanding of the analytic structure of elliptic Feynman integrals have allowed for a number of massive two-loop
Feynman integrals and scattering amplitudes to be computed~\cite{Badger:2024fgb,Becchetti:2025qlu,Becchetti:2025oyb,Badger:2025ljy,Coro:2025vgn,Becchetti:2026yxl,Buccioni:2026zva,Pozzoli:2026eiu}. 
For diphoton production specifically, the two-loop amplitudes with a closed top-quark 
loop have been computed for $q\bar{q}\to\gamma\gamma$~\cite{Becchetti:2023wev,Becchetti:2025rrz,Ahmed:2025osb} 
and $gg\to\gamma\gamma$~\cite{Maltoni:2018zvp,Becchetti:2025rrz,Ahmed:2025osb} channels. Some of these calculations have 
employed a canonical basis of elliptic master integrals~\cite{Becchetti:2025rrz,Ahmed:2024tsg}.
The results are presented as a numerical grid~\cite{Becchetti:2023wev} or analytic forms employing expansions in different kinematic regions~\cite{Becchetti:2025rrz}.
A more recent effort has been made in Ref.~\cite{Abdul-Hameed:2026aja}, where a rational basis for the
elliptic master integrals, inspired by the algebraic structure of the underlying elliptic curve~\cite{Chaubey:2026egr},
is evaluated by solving its differential equations numerically with \textsc{DiffExp}~\cite{Hidding:2020ytt},
with the aim of building an interpolation grid for the finite remainders of Ref.~\cite{Ahmed:2025osb}.
Despite these developments, an analytic representation of the finite remainders valid across 
the full kinematic range and their fast numerical evaluation over the entire physical phase space are not yet available.

The goal of this work is to derive the analytic form of the two-loop finite remainders, expressed in terms of a set of special functions,
for diphoton production in the presence of a single top-quark loop in both quark- and gluon-initiated channels,
while retaining a noncanonical basis for the elliptic sector,
which allows for efficient numerical implementation that can be used directly in the cross section evaluations.
To achieve this we introduce an approach where the pole-contributing components of the elliptic master integrals are expressed in terms
of polylogarithmic components by exploiting finite Feynman integrals built using a combination of local numerator insertions, following
Ref.~\cite{Ma:2025mog}, and soft+collinear counterterms.
With the noncanonical basis of master integrals for the elliptic sector in use, together with the canonical basis for the polylogarithmic sectors,
both the corresponding differential equation matrices and the kinematic coefficients of the amplitude are algebraic, i.e.\ free of elliptic periods.
We utilise a numerical approach to derive an independent set of polylogarithmic master integral components.
The elliptic master integral components contributing to the finite part of the amplitude, on the other hand, 
are taken as independent basis functions; hence, potential relations among them as well as
to the polylogarithmic sectors are not taken into account. 
Our approach is closely related to the one developed for the two-loop $t\bar{t}$+jet
amplitudes~\cite{Badger:2024dxo,Badger:2025ljy} and subsequently applied to $t\bar{t}W$
production~\cite{Becchetti:2026yxl}, where the noncanonical master integrals are chosen such that
their differential equations are at most quadratic in $\eps$ and the non-dlog contributions enter
only at $\mathcal{O}(\eps^4)$, i.e.\ only in the finite remainder of the two-loop amplitude.
Here we do not impose this
requirement: the elliptic master integrals  also contribute to the poles of
the amplitude, and the corresponding components are related to polylogarithmic ones by means of
finite integrals, which applies to an arbitrary choice of noncanonical elliptic master integrals.

We construct differential equations (DEs) for the special functions and solve them numerically, using both the generalised series expansion approach~\cite{Moriello:2019yhu}
and the Bulirsch--Stoer method (see, for example, the discussion in Ref.~\cite{Hairer}).
The numerical integration using the Bulirsch--Stoer method has been applied to Feynman integrals in Refs.~\cite{Boughezal:2006px,Czakon:2008zk,Czakon:2020vql,PetitRosas:2025xhm} 
and to the special functions of two-loop amplitudes involving top quarks in Refs.~\cite{Badger:2025ljy,Bera:2026qxx}. 
These numerical evaluation strategies, together with 
the analytic expressions for the finite remainders, are implemented in dedicated \textsc{Mathematica} and \textsc{C++} numerical evaluation routines.
The \textsc{C++} implementation, in particular, is suitable for direct use in Monte Carlo integration. 

The paper is organised as follows. \cref{sec:helamp} discusses the structure of the helicity amplitudes for diphoton production, 
and \cref{sec:specialfunctions} covers the construction of the special function basis. \cref{sec:implementation} describes our numerical evaluation framework, 
and \cref{sec:results} presents benchmark numerical evaluations and discusses their performance and stability. 
We give our conclusions in \cref{sec:conclusions}.

\section{Helicity amplitudes}
\label{sec:helamp}

We consider two-loop QCD corrections in the presence of a top-quark loop to the following processes that contribute to diphoton production
at the LHC
\begin{subequations}
\begin{align}
	0 \to \bar{q}(p_1) + q(p_2) + \gamma(p_3) + \gamma(p_4) \,, \\
	0 \to g(p_1) + g(p_2) + \gamma(p_3) + \gamma(p_4) \,,
\end{align}
\end{subequations}
where an all-outgoing momenta convention has been used and the following momentum conservation and on-shell
conditions are satisfied
\begin{equation}
	p_1 + p_2 + p_3 + p_4 = 0\,,\qquad p_1^2 = p_2^2 = p_3^2 = p_4^2 = 0\,.
\end{equation}
We use the scalar invariants $\vec{d} = \lbrace d_{12},d_{23}\rbrace$, where $d_{ij} = p_i \cdot p_j$, to describe the massless $2\to 2$ scattering
kinematics, together with the top-quark mass $m_t$. The physical region is defined by $0 <-d_{23} < d_{12}$. 
The partonic centre-of-mass energy is denoted by $\sqrt{s}$, where $s = 2 d_{12}$.

The $L$-loop colour decomposition of the bare helicity amplitudes is given by
\begin{subequations}
\label{eq:colourdecomposition}
\begin{align}
	\cM^{(L)} \big(1^{h_1}_{\qb},2^{h_2}_q,3^{h_3}_\gamma,4^{h_4}_\gamma\big) & = \bigg[ (4\pi)^{\eps} e^{-\eps  \gamma_E} \frac{\bar\alpha_s}{4\pi}\bigg]^L \,
	\delta^{\;\bar{i}_1}_{i_2} \, e^2 \, \cA^{(L)} \big(1^{h_1}_{\qb},2^{h_2}_q,3^{h_3}_\gamma,4^{h_4}_\gamma\big) \,, \label{eq:colourdecompositionQQ} \\
	\cM^{(L)} \big(1^{h_1}_{g},2^{h_2}_g,3^{h_3}_\gamma,4^{h_4}_\gamma\big) & = \bigg[ (4\pi)^{\eps} e^{-\eps  \gamma_E} \frac{\bar\alpha_s}{4\pi}\bigg]^L \,
	\delta^{ab} \, e^2 \, \cA^{(L)} \big(1^{h_1}_{g},2^{h_2}_g,3^{h_3}_\gamma,4^{h_4}_\gamma\big) \,, \label{eq:colourdecompositionGG}
\end{align}
\end{subequations}
where $\bar\alpha_s$ is the bare strong coupling constant and $e$ is the unit electric charge.
$\bar{i}_1$ and $i_2$ denote the anti-fundamental and fundamental colour indices of the antiquark and quark, respectively, 
while $a$ and $b$ are the adjoint colour indices of the gluons.
We work in the 't~Hooft--Veltman (tHV) scheme, in which the external momenta are kept strictly in four dimensions, while the internal ones are treated in $d = 4 - 2\eps$ dimensions.
We note that the tree-level amplitude for the $\ggaa$ process in \cref{eq:colourdecompositionGG} vanishes.

\begin{figure}[t]
  \centering
  \includegraphics[width=0.85\linewidth]{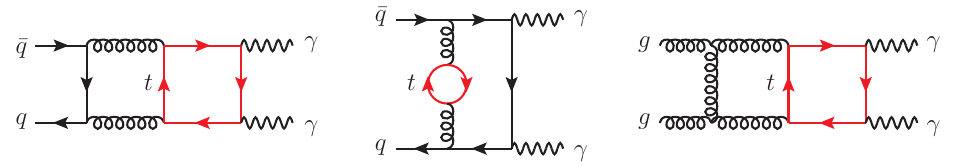} \\
	\vspace{0.2cm}\hspace{1cm}
	$A_{\qq,t}^{(2)}$ \hspace{3cm}
	$A_{\qq,n_t}^{(2)}$ \hspace{2.5cm}
	$A_{\gg,N_c}^{(2)},A_{\gg,1/N_c}^{(2)}$ 
	\caption{Representative Feynman diagrams for the two-loop partial amplitudes 
	contributing to the $pp\to \gamma\gamma$ process involving a top-quark loop.}
  \label{fig:FeynmanDiagrams}
\end{figure}

The two-loop partial amplitudes taking into account contributions only from a closed top-quark loop are
\begin{subequations}
\label{eq:partialamps}
\begin{align}
	\cA^{(2)} \big(1^{h_1}_{\qb},2^{h_2}_q,3^{h_3}_\gamma,4^{h_4}_\gamma\big) & = \bigg(N_c -\frac{1}{N_c}\bigg) \bigg[ 
	Q_t^2 \, A_{\qq,t}^{(2)\vec{h}} + n_t \, Q_q^2 \, A_{\qq,n_t}^{(2)\vec{h}} \bigg] \,, \\
	\cA^{(2)} \big(1^{h_1}_g,2^{h_2}_g,3^{h_3}_\gamma,4^{h_4}_\gamma\big) & = Q_t^2 N_c \, A_{\gg,N_c}^{(2)\vec{h}} 
	+ \frac{Q_t^2}{N_c} \, A_{\gg,1/N_c}^{(2)\vec{h}} \,,
\end{align}
\end{subequations}
where $Q_t$ and $Q_q$ are the fractional charges of the top quark and the massless initial-state quark, respectively, and
$\vec{h} = \lbrace h_1,h_2,h_3,h_4\rbrace$ denotes the helicity configuration.
Here $n_t$ is the number of heavy-quark flavours. 
In $A^{(2)\vec{h}}_{\qq,t}$, $A^{(2)\vec{h}}_{\gg,N_c}$ and $A^{(2)\vec{h}}_{\gg,1/N_c}$ both photons are attached to the top-quark loop while in
$A^{(2)\vec{h}}_{\qq,n_t}$ the two photons are coupled to the external massless quark line, see \cref{fig:FeynmanDiagrams}.

Following Refs.~\cite{Becchetti:2025rrz,Ahmed:2025osb}, we factor out the spinor phase factors $\cS^{\vec{h}}_{x}$ from the partial amplitudes to obtain the $L$-loop helicity coefficients $\helccs^{(L)\vec{h}}_{x,y}$,
\begin{equation}
        A^{(L)\vec{h}}_{x,y} = \cS^{\vec{h}}_{x} \; \helccs^{(L)\vec{h}}_{x,y} \,,
\end{equation}
where
\begin{equation}
        x \in \{\qq, \gg\} \,, \qquad
        y \in
        \begin{cases}
                \{t, n_t\} & \text{for } x = \qq \,, \\[1ex]
                \{N_c, 1/N_c\} & \text{for } x = \gg \,.
        \end{cases}
\label{eq:xylabels}
\end{equation}
Throughout this paper, we use the labels $x$ and $y$ to denote generic helicity coefficients.
The independent helicity configurations for the quark- and gluon-initiated processes are chosen as follows
\begin{subequations}
\label{eq:indephels}
\begin{align}
	\qqaa & : \quad \lbrace +-++, +-+-, +--+, +--- \rbrace \,, \\ 
	\ggaa & : \quad \lbrace ++++, -+++, +-++, ++-+, \nonumber \\
	      & \qquad +++-, --++, -+-+, +--+\rbrace \,.  
\end{align}
\end{subequations}
The spinor phase factors for the independent helicity configurations in the $\qqaa$ channel are~\cite{Becchetti:2025rrz,Ahmed:2025osb}
\begin{equation}
\begin{alignedat}{2}
	& \cS^{+-++}_{\qq}  = \frac{2\spB{3}{4}^2}{\spA{1}{3}\spB{2}{3}} \,, \qquad \qquad \qquad
	&&\cS^{+-+-}_{\qq}  = \frac{2\spA{2}{4}\spB{1}{3}}{\spA{2}{3}\spB{2}{4}} \,,  \\
	& \cS^{+--+}_{\qq}  = \frac{2\spA{2}{3}\spB{1}{4}}{\spA{2}{4}\spB{2}{3}} \,, 
	&&\cS^{+---}_{\qq}  = \frac{2\spA{3}{4}^2}{\spA{3}{1}\spB{2}{3}} \,,
\end{alignedat}
\end{equation}
while for the $\ggaa$ channel~\cite{Bern:2001df,Becchetti:2025rrz,Ahmed:2025osb}
\begin{equation}
\begin{alignedat}{2}
        & \cS^{++++}_{\gg} = \frac{\spB{1}{2}\spB{3}{4}}{\spA{1}{2}\spA{3}{4}} \,, \qquad \qquad \qquad
        &&\cS^{-+++}_{\gg} = \frac{\spA{1}{2}\spA{1}{4}\spB{2}{4}}{\spA{3}{4}\spA{2}{3}\spA{2}{4}} \,, \\
        & \cS^{+-++}_{\gg} = \frac{\spA{2}{1}\spA{2}{4}\spB{1}{4}}{\spA{3}{4}\spA{1}{3}\spA{1}{4}} \,,
        &&\cS^{++-+}_{\gg} = \frac{\spA{3}{2}\spA{3}{4}\spB{2}{4}}{\spA{1}{4}\spA{2}{1}\spA{2}{4}} \,, \\
        & \cS^{+++-}_{\gg} = \frac{\spA{4}{2}\spA{4}{3}\spB{2}{3}}{\spA{1}{3}\spA{2}{1}\spA{2}{3}} \,,
        &&\cS^{--++}_{\gg} = \frac{\spA{1}{2}\spB{3}{4}}{\spB{1}{2}\spA{3}{4}} \,, \\
        & \cS^{-+-+}_{\gg} = \frac{\spA{1}{3}\spB{2}{4}}{\spB{1}{3}\spA{2}{4}} \,,
        &&\cS^{+--+}_{\gg} = \frac{\spA{2}{3}\spB{1}{4}}{\spB{2}{3}\spA{1}{4}} \,.
\end{alignedat}
\end{equation}

The finite remainders of the two-loop helicity coefficients are obtained by subtracting the ultraviolet (UV) and infrared (IR) singularities from their bare counterparts,
\begin{equation}
        \helccfin^{(2)\vec{h}}_{x,y} =  \helccs^{(2)\vec{h}}_{x,y} - \helccpole^{(2)\vec{h}}_{x,y}  \,, 
\label{eq:finhelcoeffs}
\end{equation}
where $\helccpole^{(2)\vec{h}}_{\qq/\gg}$ contains both the UV and IR counterterms and $(x,y)$ label different components of the partial amplitude decomposition as specified
in \cref{eq:xylabels}.
The UV counterterms are obtained by renormalising the strong coupling $\as$, the external wave functions and the top-quark mass, while the IR counterterms are constructed from the universal behaviour of QCD amplitudes in the soft and collinear limits~\cite{Catani:1998twoloopIR,Becher:2009IRPRL,Becher:2009IRstructure,Gardi:2009softanomdim,Gardi:2009IRreview}.
The UV and IR counterterms needed to build $\helccpole^{(2)\vec{h}}_{\qq/\gg}$ are given in \cref{app:UVIRpoles}.
The colour-dressed finite remainders admit the same decomposition into partial finite remainders as in \cref{eq:colourdecomposition},
\begin{subequations}
\label{eq:colourdecompositionfin}
\begin{align}
        \cR^{(L)} \big(1^{h_1}_{\qb},2^{h_2}_q,3^{h_3}_\gamma,4^{h_4}_\gamma\big) & = \bigg[ (4\pi)^{\eps} e^{-\eps  \gamma_E} \frac{\alpha_s}{4\pi}\bigg]^L \,
        \delta^{\;\bar{i}_1}_{i_2} \, e^2 \, \cF^{(L)} \big(1^{h_1}_{\qb},2^{h_2}_q,3^{h_3}_\gamma,4^{h_4}_\gamma\big) \,, \label{eq:colourdecompositionfinQQ} \\
        \cR^{(L)} \big(1^{h_1}_{g},2^{h_2}_g,3^{h_3}_\gamma,4^{h_4}_\gamma\big) & = \bigg[ (4\pi)^{\eps} e^{-\eps  \gamma_E} \frac{\alpha_s}{4\pi}\bigg]^L \,
        \delta^{ab} \, e^2 \, \cF^{(L)} \big(1^{h_1}_{g},2^{h_2}_g,3^{h_3}_\gamma,4^{h_4}_\gamma\big) \,, \label{eq:colourdecompositionfinGG}
\end{align}
\end{subequations}
where $\as$ is the renormalised strong coupling constant. 
The partial finite remainders, restricted to contributions involving one top-quark loop, read
\begin{subequations}
\label{eq:partialfinrem}
\begin{align}
        \cF^{(2)} \big(1^{h_1}_{\qb},2^{h_2}_q,3^{h_3}_\gamma,4^{h_4}_\gamma\big) & = \bigg(N_c -\frac{1}{N_c}\bigg) \bigg[
        Q_t^2 \, F_{\qq,t}^{(2)\vec{h}} + n_t\,Q_q^2 \, F_{\qq,n_t}^{(2)\vec{h}} \bigg] \,, \\
        \cF^{(2)} \big(1^{h_1}_g,2^{h_2}_g,3^{h_3}_\gamma,4^{h_4}_\gamma\big) & = Q_t^2 N_c \, F_{\gg,N_c}^{(2)\vec{h}}
        + \frac{Q_t^2}{N_c} \, F_{\gg,1/N_c}^{(2)\vec{h}} \,.
\end{align}
\end{subequations}
The finite remainder of the helicity coefficients defined in \cref{eq:finhelcoeffs} is related to the partial finite remainders by
\begin{equation}
	F^{(L)\vec{h}}_{x,y} = \cS^{\vec{h}}_{x} \; \tilde{F}^{(L)\vec{h}}_{x,y} \,.
\end{equation}

The derivation of the helicity finite remainder coefficients as well as their numerical evaluation 
(including the evaluation of the corresponding special functions) is done with the renormalisation scale $\mu$ and
the top-quark mass, $m_t$, set to one. The actual values of both $\mu$ and $m_t$ are recovered by exploiting the explicit scale dependence
of the finite remainders. 
For the quark-initiated channel the scale dependence of the helicity finite remainder coefficients is
\begin{subequations}
\label{eq:mudepQQ}
\begin{align}
	\tilde{F}^{(2)\vec{h}}_{\qq,t}(\mu) = & \; \tilde{F}^{(2)\vec{h}}_{\qq,t}(\mu=1) \,, \\
	\tilde{F}^{(2)\vec{h}}_{\qq,n_t}(\mu) = & \; \tilde{F}^{(2)\vec{h}}_{\qq,n_t}(\mu=1) 
	- \frac{2}{3} \log(\mu^2) \; \tilde{F}^{(1)\vec{h}}_{\qq,q}(\mu=1) \nonumber \\
	& + \tilde{F}^{(0)\vec{h}}_{\qq,q} \bigg\lbrace \frac{1}{3} \log^3(\mu^2) + \left[1 - \frac{2}{3}\log(-2d_{12})\right] \log^2(\mu^2) 
	  \bigg\rbrace\,,
\end{align}
\end{subequations}
while for the gluon-initiated process
\begin{subequations}
\label{eq:mudepGG}
\begin{align}
	\tilde{F}^{(2)\vec{h}}_{\gg,N_c}(\mu)   = & \; \tilde{F}^{(2)\vec{h}}_{\gg,N_c}(\mu=1) 
	                                               + \bigg[ 2\log(\mu^2)\log(-2d_{12}) - \log^2(\mu^2) \bigg]\tilde{F}^{(1)\vec{h}}_{\gg,t} \,, \\
	\tilde{F}^{(2)\vec{h}}_{\gg,1/N_c}(\mu) = & \; \tilde{F}^{(2)\vec{h}}_{\gg,1/N_c}(\mu=1) \,.
\end{align}
\end{subequations}
The tree-level and one-loop finite remainders appearing in \cref{eq:mudepQQ,eq:mudepGG} take the following form, analogous to \cref{eq:partialfinrem}
\begin{subequations}
\begin{align}
	\cF^{(0)}\big(1^{h_1}_{\qb},2^{h_2}_q,3^{h_3}_\gamma,4^{h_4}_\gamma\big) & = Q_q^2 \; F^{(0)\vec{h}}_{\qq,q} \,, \\
	\cF^{(1)}\big(1^{h_1}_{\qb},2^{h_2}_q,3^{h_3}_\gamma,4^{h_4}_\gamma\big) & = \bigg(N_c -\frac{1}{N_c}\bigg) \; Q_q^2 \; F^{(1)\vec{h}}_{\qq,q} \,, \\
	\cF^{(1)}\big(1^{h_1}_g,2^{h_2}_g,3^{h_3}_\gamma,4^{h_4}_\gamma\big) & = Q_t^2 \; F^{(1)\vec{h}}_{\gg,t} \,, \label{eq:oneloopdecompositionGG}
\end{align}
\end{subequations}
where in \cref{eq:oneloopdecompositionGG} we retain only the contribution involving a top-quark loop.
We define the hard functions for the $\qqaa$ and $\ggaa$ processes as follows
\begin{subequations}
\label{eq:hardfunctions}
\begin{align}
	H^{(0)}_{\qq,q} &= 2 N_c \, \sum_{\vec{h}} \big( F^{(0)\vec{h}}_{\qq,q} \big)^* F^{(0)\vec{h}}_{\qq,q}  \,, \\
	H^{(1)}_{\gg,t} &= 2 (N_c^2-1) \, \sum_{\vec{h}} \big( F^{(1)\vec{h}}_{\gg,t} \big)^* F^{(1)\vec{h}}_{\gg,t}  \,, \\
	H^{(2)}_{\qq,t} &= 2 N_c \bigg(N_c - \frac{1}{N_c}\bigg) \, \mathrm{Re} \,\sum_{\vec{h}} \big( F^{(0)\vec{h}}_{\qq,q} \big)^* F^{(2)\vec{h}}_{\qq,t}  \,, \\
	H^{(2)}_{\qq,n_t} &= 2 N_c \bigg(N_c - \frac{1}{N_c}\bigg) \, \mathrm{Re} \, \sum_{\vec{h}} \big( F^{(0)\vec{h}}_{\qq,q} \big)^* F^{(2)\vec{h}}_{\qq,n_t}  \,, \\
	H^{(2)}_{\gg,t} &= 2 (N_c^2-1) \, \mathrm{Re} \,\sum_{\vec{h}} \big( F^{(1)\vec{h}}_{\gg,t} \big)^* 
			   \bigg[ N_c F^{(2)\vec{h}}_{\gg,N_c} + \frac{1}{N_c} F^{(2)\vec{h}}_{\gg,1/N_c}\bigg]  \,, 
\end{align}
\end{subequations}
where the sum over colours has been explicitly performed and $\vec{h}$ refers to the independent helicity configurations given in \cref{eq:indephels}.
The factor of 2 accounts for the parity-conjugated helicity configurations.

The computation of the helicity coefficients follows a diagrammatic approach, 
in which the relevant set of Feynman diagrams is generated using \textsc{Qgraf}~\cite{Nogueira:1991qform}. 
We then perform the colour decomposition and extract the bare partial amplitudes appearing in \cref{eq:partialamps}.
These partial amplitudes are projected onto helicity coefficients employing the four-dimensional tensor decomposition~\cite{Peraro:2019cjj,Peraro:2020sfm}.
Such projections are discussed in detail in Section 2 of Ref.~\cite{Becchetti:2025rrz}, and for conciseness we do not repeat the construction here.
We make use of a combination of in-house \textsc{Form}~\cite{Ruijl:2017form42} and \textsc{Mathematica} scripts to carry out colour decomposition, helicity coefficient projection
and the relevant $\gamma$-matrix algebra.
The bare two-loop helicity coefficients are then expressed in terms of scalar Feynman integrals $I_i$ as
\begin{equation}
	\helccs^{(2)\vec{h}}_{x,y} = \sum_i c^{(2)\vec{h}}_{x,y;i}(\eps,\vec{d}) \; I_i \,, 
	\label{eq:unreducedbare}
\end{equation}
where we organise $I_i$ according to the distinct Feynman integral families. Each Feynman diagram can be mapped to at least one of the Feynman integral families.
The scalar Feynman integrals are then reduced to a basis of master integrals $\cJ_i$ by means of integration-by-parts (IBP) reduction.
We generate the IBP relations using the \textsc{NeatIBP} package~\cite{Wu:2023upw} and solve them numerically over finite fields within the \textsc{FiniteFlow} framework~\cite{Peraro:2019svx}.
After the IBP reduction, the bare two-loop helicity coefficients can be written as linear combinations of master integrals $\cJ_i$
\begin{equation}
	\helccs^{(2)\vec{h}}_{x,y} = \sum_{i=1}^{169} \hat{c}^{(2)\vec{h}}_{x,y;i}(\eps,\vec{d}) \; \cJ_i \,.
\end{equation}
There are 169 master integrals covering all partial amplitudes appearing on the RHS of \cref{eq:partialamps}.
Expressing master integrals in terms of  special function monomials $\mathrm{mon}_i(\sfvec)$ and performing a Laurent expansion in $\eps$ we obtain
\begin{equation}
	\helccs^{(2)\vec{h}}_{x,y} = \sum_{i} \sum_{j=-3}^{0} \eps^j e^{(2)\vec{h}}_{x,y;ij}(\vec{d}) \; \mathrm{mon}_i(\sfvec) + \cO(\eps) \,.
\end{equation}
Here the deepest $\eps$ pole varies between partial amplitudes and helicity configurations.
The choice of master integrals and the construction of the special function basis are discussed in detail in \cref{sec:specialfunctions}.
Expressing the UV and IR pole terms in \cref{eq:finhelcoeffs} in the same special function basis as for the bare helicity coefficients, the finite remainder
takes the following form
\begin{equation}
	\tilde{F}^{(2)\vec{h}}_{x,y} = \sum_{i}  r^{(2)\vec{h}}_{x,y;i}(\vec{d}) \; \mathrm{mon}_i(\sfvec) + \cO(\eps) \,.
\label{eq:finremSF}
\end{equation}
In this work we employ the finite-field method~\cite{Peraro:2016wsq} where all the algebraic operations needed to derive the finite remainder in \cref{eq:finremSF} from
the unreduced bare helicity coefficient in \cref{eq:unreducedbare} are replaced by numerical evaluation over finite fields. Essentially, we set up a framework
to evaluate the rational coefficients of the special function monomials in the helicity finite remainder $\tilde{F}^{(2)\vec{h}}_{x,y}$ numerically over finite fields.
The analytic form of $\tilde{F}^{(2)\vec{h}}_{x,y}$ is then reconstructed from multiple numerical evaluations, employing the built-in analytic reconstruction
machinery implemented in \textsc{FiniteFlow}.

\section{Special functions}
\label{sec:specialfunctions}

\begin{figure}[t]
  \centering
  \includegraphics[width=0.95\linewidth]{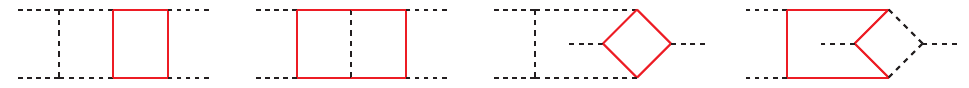} \\
	PL1 \hspace{2.6cm}
	PL2 \hspace{2.2cm}
	NPL1 \hspace{2.6cm}
	NPL2 
	\caption{Two-loop master integral families for the $pp\to\gamma\gamma$ process involving a top-quark loop.
	The dashed black (solid red) lines represent massless (massive) particles.}
  \label{fig:MIfamilies}
\end{figure}

We first discuss the master-integral basis used in our calculation. The integral families comprise two planar and two nonplanar families, as shown in \cref{fig:MIfamilies}.
To cover the $\qqaa$ and $\ggaa$ amplitudes, we need to apply permutations of the external momenta to the PL1, PL2 and NPL2 families.
We then obtain the minimal set of master integrals across all families by deriving the corresponding mapping rules with \textsc{LiteRed}~\cite{Lee:2012cn,Lee:2013mka}, which results in 169 master integrals.
The differential equations are constructed for the full system of 169 master integrals. 
For the PL1, PL2, NPL1 and NPL2 families, we employ the bases of master integrals taken from Refs.~\cite{Caron-Huot:2014lda,Becchetti:2017abb,Xu:2018eos,AH:2023ewe,Ahmed:2024tsg,Becchetti:2025rrz}, except for the four top-sector master integrals of the NPL1 family. 
We denote the top-sector master integrals of NPL1 by $\lbrace \cJ^{\nplone}_{33}, \cJ^{\nplone}_{34}, \cJ^{\nplone}_{35}, \cJ^{\nplone}_{36}\rbrace$,
and they are chosen as
\begin{subequations}
\label{eq:MIsNPL1top}
\begin{align}
	\cJ^{\nplone}_{33} &= \eps^4 (1+2\eps) I^{\nplone}_{111111100} \,, \\
	\cJ^{\nplone}_{34} &= \eps^4 (1+2\eps) I^{\nplone}_{11111110-1} \,, \\
	\cJ^{\nplone}_{35} &= \frac{\eps^4 (1+2\eps)}{d_{12} + 2 d_{23}} \bigg[ 
				(1-3\eps) I^{\nplone}_{1111111-1-1} - (1-2\eps) I^{\nplone}_{11111110-2} \nonumber \\
			& \quad - 2 d_{23}(1-3\eps) I^{\nplone}_{11111110-1} + 2(d_{23}-4) d_{23} \,\eps\, I^{\nplone}_{111111100} \bigg] \\
			& \quad +\; \mathrm{subsector\;\,integrals} \,, \nonumber \\
	\cJ^{\nplone}_{36} &= \eps^4 (1+2\eps) I^{\nplone}_{1111111-1-1} \,,
\end{align}
\end{subequations}
where we set $m_t^2=1$ and  
\begin{equation}
	I^{\nplone}_{a_1 a_2 a_3 a_4 a_5 a_6 a_7 a_8 a_9} = e^{2\eps\gamma_E} \int \frac{d^d k_1}{i \pi^{d/2}} \frac{d^d k_2}{i \pi^{d/2}}
	\frac{D_8^{-a_8} D_9^{-a_9}}{ D_1^{a_1} D_2^{a_2} D_3^{a_3} D_4^{a_4} D_5^{a_5} D_6^{a_6} D_7^{a_7} } \,.
\end{equation}
The inverse propagators are
\begin{align}
\label{eq:NPL1propagators}
  D_1&=k_1^2, & D_2&=(k_1-p_1)^2, & D_3&=(k_1-p_1-p_2)^2,\nonumber\\
  D_4&=k_2^2-m_t^2, & D_5&=(k_2+p_1+p_2+p_3)^2-m_t^2, & D_6&=(k_1+k_2)^2-m_t^2,\nonumber\\
  D_7&=(k_1+k_2+p_3)^2-m_t^2, & D_8&=(k_1+p_4)^2, & D_9&=(k_2-p_1)^2-m_t^2,
\end{align}
where $D_8,D_9$ are irreducible scalar products (ISPs), so their indices are always $\le 0$.
The top-sector integrals of the NPL1 family  are chosen such that the resulting differential equations are free of (i) factors of $\eps$ in the denominators and (ii) spurious singularities in the physical scattering region. In particular, the connection matrices of the full system of differential equations are at most quartic in $\eps$.
When implementing the canonical basis of master integrals in the finite-field framework, we strip the overall square-root factors from the relevant integrals and restore them in the final results.
The complete definitions of the master integrals, together with the corresponding differential equations, are provided in the ancillary files accompanying this paper.

To construct a set of special functions, we expand the master integrals, normalised to start at $\eps^0$, up to order $\eps^4$,
\begin{equation}
        \cJ_i = \sum_{w=0}^{4} \eps^w\, \cJ_{iw} + \cO(\eps^5) \,,
        \label{eq:MIexpansion}
\end{equation}
and refer to $\cJ_{iw}$ as the weight-$w$ component of $\cJ_i$.
We have verified that the master-integral coefficients in the bare helicity coefficients have poles no deeper than $1/\eps^4$, so higher-order terms in \cref{eq:MIexpansion} are not needed.
The 169 two-loop master integrals,
together with the 15 one-loop master integrals that also enter the finite remainders (ten for
$\ggaa$ and five for $\qqaa$), give $4\times184=736$ components of weight one to four,
since the weight-zero components are rational numbers. These components are not independent:
many of them are constants, and many others can be written as polynomials in components of
equal or lower weight. The coefficients of these polynomials are rational numbers, possibly
multiplied by transcendental constants such as $\pi^2$ and $\zeta_3$; in the elliptic sector
they can also be algebraic functions of the kinematics. We call a minimal set of components
in terms of which all others that are present in the finite remainder can be expressed in this way the \textit{special functions}.
Bases of this kind are well known from two-loop five-point amplitudes, where they are called
\textit{pentagon functions}~\cite{Gehrmann:2018yef,Chicherin:2020oor,Chicherin:2021dyp,Abreu:2023rco}, and they have since been constructed for several other
processes. Here we keep the generic name special functions.

We treat the master integrals in two classes. All master integrals except the six of the elliptic
sector NPL1 are normalised such that their differential equations are in canonical form and do not
involve the elliptic ones. Together with the one-loop master integrals, these are 178
polylogarithmic master integrals. Their relations have rational coefficients and are found
numerically in \cref{sec:polylog}. For the six elliptic
master integrals of NPL1, the differential equations are not in canonical form and the
relations in general have kinematics-dependent coefficients. In \cref{sec:elliptic} we express their
pole-contributing components through polylogarithmic ones, using a basis of finite integrals.

\subsection{Construction of the special function basis in the polylogarithmic sector} 
\label{sec:polylog}

The standard way to find relations among the polylogarithmic master integral components, used in the construction of pentagon functions and
their generalisations~\cite{Gehrmann:2018yef,Chicherin:2020oor,Chicherin:2021dyp,Abreu:2023rco}, is based on their
symbols~\cite{Goncharov:2010jf,Duhr:2011zq,Duhr:2012fh}. Alternatively, relations can be searched for
numerically with an integer-relation algorithm such as PSLQ~\cite{Ferguson:1999pslq,Bailey:1999nv},
applied to the values of the components and of their products at a phase-space point. This
requires the components to very high precision, which is computationally expensive, and it
cannot guarantee that all relations have been found. In this work we follow a different
numerical approach, which is described in the following.

Setting $m_t^2=1$ and collecting the polylogarithmic master integrals into 
a vector $\vec{\cJ}$, we write the differential equations as
\begin{equation} \label{eq:canonicalform}
        \partial_x \vec{\cJ}=\eps\,A_x(d_{12},d_{23})\,\vec{\cJ},
        \qquad x\in\{d_{12},d_{23}\},
\end{equation}
where the entries of $A_x$ are rational functions and square roots of the invariants. The
rows of the polylogarithmic master integrals do not involve the elliptic ones, so this subset is
closed under differentiation and can be treated on its own. Inserting the expansion
\cref{eq:MIexpansion} turns the differential equations into a recursion in the weight,
\begin{equation}
	\partial_x \cJ_{iw}=\sum_j (A_x)_{ij}\,\cJ_{j,w-1},
	\qquad \cJ_{j,-1}=0 .
	\label{eq:compDE}
\end{equation}
The weight-$0$ components are rational numbers, and the derivative of a weight-$w$ component
is a combination of weight-$(w-1)$ components with algebraic coefficients.
\Cref{eq:compDE} and the numerical values of all components at one phase-space point (referred to as the reference point below)
are the only input of the method. 

The method rests on the fact that two functions with the same gradient differ by a
constant. Assume that the independent functions of the lower weights, i.e.\ the special functions
$S_v=\{\sfn{v}{j}\}$ of weight $v=1,\dots,w-1$, are known, together with the relations that express every other component of weight $<w$ in
terms of them. At weight $w$ we consider the ansatz
\begin{equation}
	\Phi=\sum_j a_j\,\phi_j ,\qquad a_j\in\mathbb Q ,
	\label{eq:ansatz}
\end{equation}
where the $\phi_j$ run over (i) all weight-$w$ components, (ii) all products of lower-weight
basis functions of total weight $w$ (at $w=4$: $S_3S_1$, $S_2S_2$, $S_2S_1^2$, $S_1^4$), and
(iii) $\pi^2$ and $\zeta_3$ times the products of weight $w-2$ and $w-3$. Pure constants
are left out, since their derivative vanishes. Using \cref{eq:compDE}, the Leibniz rule
and the relations of lower weight, the derivative of every $\phi_j$ is brought to the form
\begin{equation}
	\partial_x \phi_j=\sum_k R^{(x)}_{kj}(d_{12},d_{23})\;g_k ,
	\label{eq:Rmatrix}
\end{equation}
where the $g_k$ are the monomials of weight $w-1$ in the basis functions (with $\pi^2$ and
$\zeta_3$ counted as monomials of weight $2$ and $3$), and the $R^{(x)}_{kj}$ are algebraic
functions of the invariants. 
As the $g_k$ are linearly independent, $\Phi$ is a
constant if and only if
\begin{equation}
	\sum_j R^{(x)}_{kj}(d_{12},d_{23})\;a_j=0
	\qquad\text{for all } k,\ x \text{ and all } (d_{12},d_{23}).
	\label{eq:nullcond}
\end{equation}
Evaluating $R^{(x)}_{kj}$ at a few random rational points gives a numerical matrix $\mathbf N$, with
one column per element of the ansatz and one row per invariant, point and monomial $g_k$.
The relations are the null vectors of $\mathbf N$, each determined up to an additive constant of
weight $w$.

The algorithm proceeds as follows.
The weight-$0$ components are rational numbers, which are fixed from the numerical values
of the master integrals. The relations are then constructed weight by weight,
$w=1,\dots,4$:
\begin{enumerate}
	\item Components whose reduced derivative vanishes identically are constants. They are
	fixed from the numerical values and removed from the ansatz.
	\item The remaining elements of the ansatz, i.e.\ the columns of $\mathbf N$, are ordered as
	\begin{equation*}
		\big(\,\text{two-loop components}\;\big|\;\text{one-loop components}\;\big|\;
		\text{products}\,\big),
	\end{equation*}
	where ``products'' stands for the products of lower-weight basis functions, with and
	without a factor $\pi^2$ or $\zeta_3$. This order decides which components are kept
	as independent functions, as explained below.  
	\item The matrix $\mathbf N$ is built at six random rational points, each close to the
	reference point in $d_{12}$ and $d_{23}$. Near the reference point every square root
	is evaluated on the same branch as in the numerical values. We explain the choice of six points below.
	\item For the larger matrices the null space is first computed at machine precision with
	unit-normalised columns.
	Only the columns
	involved in these null vectors are kept, and the null space is recomputed with
	high precision. This strategy is particularly convenient for larger matrices that appear in higher weights. 
	\item The null space is brought to reduced row echelon form. Each row is one relation,
	solved for its first non-zero column, and its coefficients are rationalised.
	\item Every relation is verified independently of $\mathbf N$. The derivatives of both sides are
	computed from the differential equations and reduced with the known relations. Their
	difference is a polynomial in the lower-weight basis functions and must vanish
	identically. 
	\item The additive constant is obtained by evaluating the relation at the reference point
	with numerical values from \textsc{AMFlow}~\cite{Liu:2022chg}. We find that it vanishes
	at weight $1$ and is a rational multiple of $\pi^2$, $\zeta_3$ and $\pi^4$ at weights $2$,
	$3$ and $4$, respectively. Its coefficient is fixed from the real part of the residual.
	
\end{enumerate}

The null space determines the relations only up to linear recombination, so it does not
fix which components are kept as independent functions and which are expressed through
them. This choice is made by the order of the columns of $\mathbf N$. A row of the reduced
row echelon form has the entry $1$ in its pivot column, the first column in which it is
non-zero, and all its other non-zero entries lie in non-pivot columns further to the right.
The row is therefore solved for the pivot,
\begin{equation}
	\phi_{\rm piv}=-\sum_{j>{\rm piv}}a_j\,\phi_j+\text{const},
\end{equation}
and a column becomes a pivot exactly when its function equals, up to a constant, a linear
combination of the functions in the columns to its right. The independent functions are
thus selected starting from the right end of the list, and the functions to be eliminated
are placed at the left.

We order the columns such that, wherever there is a choice, a function is expressed
through functions of lower weight, of fewer loops or with fewer propagators. As described
in step~2, the products stand at the right end, the one-loop components to their left and
the two-loop components at the left end. The one-loop components are ordered as massive
boxes, triangles, massless boxes, massive bubbles, massless bubbles and the tadpole. The
two-loop components follow our list of master integrals from the top sector to the lower sectors; among them no preference is
imposed. This order has two consequences:
\begin{itemize}	
	\item \emph{Higher weight through lower weight.} The products stand at the right end. By
	construction they are linearly independent among themselves, so no relation is solved
	for a product. A component that equals a polynomial in lower-weight functions is
	therefore always expressed through the products.
	\item \emph{Two-loop through one-loop.} The two-loop components stand to the left of the
	one-loop ones, so a relation involving both is solved for a two-loop component, and
	one-loop components are kept as independent functions whenever possible. 
	
\end{itemize}
A different order leads to a different but equivalent choice of independent functions. The
number of independent functions at each weight, and the space they span together with the
products and constants, do not depend on the order.

A relation must hold for all values of the invariants, whereas the matrix $\mathbf N$ samples the
derivatives at a finite number of points. The minimal number of points required cannot be
determined in advance, but the procedure provides its own check. A genuine relation is
satisfied at every point and is therefore always contained in the null space of $\mathbf N$. If the
number of points is too small, the null space may in addition contain spurious vectors.
These fail the verification of step~6, which is independent of $\mathbf N$. The number of points is
thus sufficient if every null vector passes this verification. We use six points, for which
this is the case at all weights.

\subsection{Relations for elliptic master integral components}
\label{sec:elliptic}

The next step is to write the master integral components belonging to the elliptic sector that contribute to the poles of the amplitudes
in terms of polylogarithmic master integral components. Such relations exist since the poles of two-loop amplitudes are solely determined by
the tree-level and one-loop amplitudes as well as the UV and IR counterterms, none of which contain elliptic functions.
Finite Feynman integrals have been constructed in a variety of ways, see e.g.\
Refs.~\cite{vonManteuffel:2014qoa,Gambuti:2023eqh,delaCruz:2024xsm,Dhani:2026cxx,Georgoudis:2026han,Figueiredo:2026ksz}.
Our strategy follows Ref.~\cite{Ma:2025mog}, where a set of finite Feynman integrals is employed to derive relations among master integral components.
This approach is implemented in the \textsc{FFI} package, in which the finite integrals are constructed either in higher dimensions, 
combined with dimensional recurrence relations, or in four dimensions through local numerator insertions.
We follow the latter approach. For the one relation that cannot be obtained in this way, we additionally construct a finite integral 
by explicitly subtracting the soft and collinear divergences~\cite{Georgoudis:2026han}, as described below.

We search for a basis of finite integrals of the NPL1 top sector which can be used to extract the relations 
among the pole-contributing elliptic master integral components and the polylogarithmic counterparts.
This is done by expressing the finite integrals in terms of the top sector master integrals in \cref{eq:MIsNPL1top},
applying \cref{eq:MIexpansion} to the master integrals and expanding the integral relations in $\eps$, such that the relations among master integral components
can be extracted order by order in $\eps$.
It is essential to ensure that if the basis of finite integrals is used in the amplitude, together with the respective coefficients, 
those integrals contribute only to the finite part of the amplitude. 
We note that in addition to $\lbrace \cJ^{\nplone}_{33},\cdots,\cJ^{\nplone}_{36} \rbrace$, two subsector integrals with six propagators 
also belong to the elliptic sector. 
Those two integrals have been chosen such that they already contribute only to the finite part
of the amplitude, i.e.\ their components vanish up to $\mathcal{O}(\eps^3)$, and the corresponding differential equations also take a simple form, i.e.\ they are quadratic in $\eps$.

To construct the candidates for finite integrals we introduce local numerator insertions to the top sector of the NPL1 family 
which vanish in the soft and collinear limits. One of the possible choices of such an integral with numerator insertion is
\begin{equation}
\label{eq:finiteIntegralLocalNum}
	G^{\nplone}_{\mathrm{cand};n} = 
	 e^{2\eps\gamma_E} \int \frac{d^d k_1}{i \pi^{d/2}} \frac{d^d k_2}{i \pi^{d/2}}
	\frac{\left[-\frac{1}{4}\mathrm{tr}(\slashed{p}_2 \slashed{p}_3 \slashed{p}_1\slashed{k}_1)\right]^n}{ 
	D_1 D_2 D_3 D_4 D_5 D_6 D_7 } \,,
\end{equation}
where $D_i$'s are specified in \cref{eq:NPL1propagators} and $n$ takes positive integer values. 
Note that $n=0$ corresponds to the $I^{\nplone}_{111111100}$ integral.
The numerator in \cref{eq:finiteIntegralLocalNum} vanishes in the soft limit, $k_1 \to p_1$, and in the collinear limits, $k_1 || p_1$ and $(k_1-p_1) || p_2$.
We can simply take $G^{\nplone}_{\mathrm{cand};1},\cdots,G^{\nplone}_{\mathrm{cand};4}$ as the four finite master integral candidates for
the NPL1 top sector, as they are finite (as checked with \textsc{AMFlow}) and independent such that they form a basis.
Expressing them in terms of master integrals in \cref{eq:MIsNPL1top} through IBP reduction, we have
\begin{align}
\label{eq:GintermsofJ}
\begin{aligned}
        \eps^4\, G^{\nplone}_{\mathrm{cand};1} &= a_{34}\, \cJ^{\nplone}_{34} + \text{subsectors} \,, \\
        \eps^4\, G^{\nplone}_{\mathrm{cand};2} &= b_{34}\, \cJ^{\nplone}_{34} + b_{36}\, \cJ^{\nplone}_{36} + \text{subsectors} \,, \\
        \eps^4\, G^{\nplone}_{\mathrm{cand};3} &= \eps\, c_{33}\, \cJ^{\nplone}_{33}
                + c_{34}\, \cJ^{\nplone}_{34} + c_{35}\, \cJ^{\nplone}_{35} + c_{36}\, \cJ^{\nplone}_{36} + \text{subsectors} \,, \\
        \eps^4\, G^{\nplone}_{\mathrm{cand};4} &= \eps\, d_{33}\, \cJ^{\nplone}_{33}
                + d_{34}\, \cJ^{\nplone}_{34} + d_{35}\, \cJ^{\nplone}_{35} + d_{36}\, \cJ^{\nplone}_{36} + \text{subsectors} \,,
\end{aligned}
\end{align}
where the coefficients $a,b,c,d$ are rational functions of the kinematics and $\eps$, finite and non-vanishing at $\eps=0$.
The subsector integrals appearing in \cref{eq:GintermsofJ} may include the two elliptic
subsector integrals mentioned above. Since their components vanish up to $\mathcal{O}(\eps^3)$,
they do not enter the relations, and the remaining subsector integrals are polylogarithmic.
The presence of the $\eps$ factor in front of $\cJ^{\nplone}_{33}$ shifts the $\cJ^{\nplone}_{33}$ master integral component contributions
such that at $\eps^i$, the $\cJ^{\nplone}_{33,i-1}$ component contributes, i.e.\ $\cJ^{\nplone}_{33,3}$ appears at order $\eps^4$ and its
relation to the polylogarithmic components cannot be extracted.
We have checked that the $\cJ^{\nplone}_{33,3}$ component contributes to the $1/\eps$ pole of the amplitude,
so it should be expressible in terms of polylogarithmic master integral components.
We further scanned a large set of other candidates for local numerator insertions, in the same spirit as \cref{eq:finiteIntegralLocalNum},
to remove such $\eps$ factors, and in every basis we tried, 
when they are expressed in terms of $\cJ^{\nplone}_{i}$, $i\in \lbrace 33, \cdots,36\rbrace$, we always find at least a factor of $\eps$ multiplying $\cJ^{\nplone}_{33}$.
Using the \textsc{FFI} package, we have verified that the relation involving $\cJ^{\nplone}_{33,3}$ is likewise not obtained.

To derive the missing relation, we make use of the constraints in the soft and collinear limits to build the finite integral.
In the soft limit, $k_1 \to p_1$, we have that
\begin{align}
	\label{eq:E6E7defs}
	D_6 & \to (k_2 + p_1)^2 - m_t^2 \equiv E_6  \,, \\
	D_7 & \to (k_2 + p_1 + p_3)^2 - m_t^2 \equiv E_7  \,, 
\end{align}
where $D_6$ and $D_7$ are defined in \cref{eq:NPL1propagators}, such that the scalar top-sector NPL1 integral $I^{\nplone}_{111111100}$
factorises into a product of one-loop integrals
\begin{align}
	I^{\nplone}_{111111100} \big|_{\mathrm{soft}} & = I\bigg(\usepix{1.5cm}{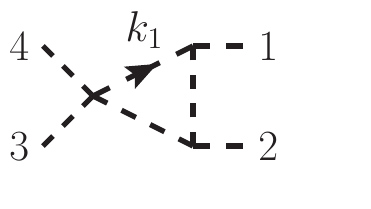}{0}{0}{40}{0}\bigg) \; 
			   			           I\bigg(\usepix{1.5cm}{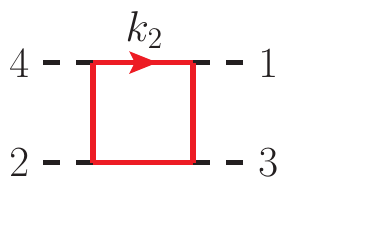}{0}{0}{40}{0}\bigg) \,,
\end{align}
where
\begin{align}
	I\bigg(\usepix{1.5cm}{figs/tri_1_2_34x000.pdf}{0}{0}{40}{0}\bigg) & = e^{\eps\gamma_E} \int \frac{d^d k_1}{i \pi^{d/2}} 
											\frac{1}{D_1 D_2 D_3 } \,, \label{eq:massless1Ltriangle} \\
	I\bigg(\usepix{1.5cm}{figs/box1324xMMMM.pdf}{0}{0}{40}{0}\bigg) & = e^{\eps\gamma_E} \int \frac{d^d k_2}{i \pi^{d/2}} \frac{1}{D_4 D_5 E_6 E_7}\,.
\end{align}
We note that the massless one-loop triangle in \cref{eq:massless1Ltriangle} can be further reduced to the massless one-loop bubble integral.
The soft-subtracted $ I^{\nplone}_{111111100}$ integral takes the form
\begin{equation}
	\label{eq:softsubtracted}
	I^{\nplone}_{111111100} - I^{\nplone}_{111111100} \big|_{\mathrm{soft}} = 
	e^{2\eps\gamma_E} \int \frac{d^d k_1}{i \pi^{d/2}} \frac{d^d k_2}{i \pi^{d/2}} 	\frac{1}{D_1 D_2 D_3 D_4 D_5} \bigg[ \frac{1}{D_6 D_7} - \frac{1}{E_6 E_7} \bigg] \,.
\end{equation}
We now construct the counterterms for the collinear regions. For the case where $k_1 || p_1$,
it is convenient to partial fraction the bracketed term in the integrand of \cref{eq:softsubtracted} as follows
\begin{equation}
	\frac{1}{D_6 D_7} - \frac{1}{E_6 E_7}  = -\frac{\delta_6}{D_6 E_6 E_7} - \frac{\delta_7}{D_6 D_7 E_7} \,, \label{eq:lineA} 
\end{equation}
where
\begin{align}
  \delta_6&=k_1^2+2k_1\!\cdot\!k_2-2k_2\!\cdot\!p_1,\\
  \delta_7&=k_1^2+2k_1\!\cdot\!k_2+2k_1\!\cdot\!p_3-2k_2\!\cdot\!p_1+2(d_{12}+d_{23}) \,.
\end{align}
At  $k_1 = x p_1$, we have that
\begin{align}
\begin{aligned}
	\delta_6&=-\frac{k_2\cdot p_1}{d_{12}} D_3 \,, \\
	\delta_7&=\frac{d_{12}+d_{23}-k_2\cdot p_1}{d_{12}} D_3  \,.
\label{eq:delta67atColl1}
\end{aligned}
\end{align}
For the second collinear region, $(k_1 -p_1)||p_2$, we use the following partial fractioned form instead
\begin{equation}
	\frac{1}{D_6 D_7} - \frac{1}{E_6 E_7}  = -\frac{\delta_6}{D_6 E_6 D_7} - \frac{\delta_7}{E_6 D_7 E_7} \,, \label{eq:lineB}
\end{equation}
and at $k_1 = p_1 + y p_2$,
\begin{align}
\begin{aligned}
	\delta_6&=\frac{d_{12}+k_2\cdot p_2}{d_{12}} D_1 \,, \\
	\delta_7&=\frac{d_{12}+d_{23}+k_2\cdot p_2}{d_{12}} D_1  \,.
\label{eq:delta67atColl2}
\end{aligned}
\end{align}
Substituting \cref{eq:delta67atColl1} into \cref{eq:lineA} and \cref{eq:delta67atColl2} into \cref{eq:lineB}
we obtain the collinear counterterms to the soft-subtracted integral.
Collecting all the ingredients together we define the soft- and collinear-subtracted $I^{\nplone}_{111111100}$ as
\begin{align}
\begin{aligned}
	\label{eq:I_NPL1_scalarsubtracted}
	\tilde{G}^{\nplone}_{\mathrm{cand};\mathrm{sc}} = & I^{\nplone}_{111111100} - I^{\nplone}_{111111100}\big|_{\mathrm{soft}}    \\
	& - e^{2\eps\gamma_E} \int \frac{d^d k_1}{i \pi^{d/2}} \frac{d^d k_2}{i \pi^{d/2}}  \bigg[
		 \frac{k_2\cdot p_1/d_{12}}{D_1 D_2 D_4 D_5 D_6 E_6 E_7}  \\
	&  + \frac{(k_2\cdot p_1 - d_{12} - d_{23})/d_{12}}{D_1 D_2 D_4 D_5 D_6 D_7 E_7} 
	   - \frac{(k_2\cdot p_2 + d_{12})/d_{12}}{D_2 D_3 D_4 D_5 D_6 E_6 D_7} \\
	&   - \frac{(k_2\cdot p_2 + d_{12}+d_{23})/d_{12}}{D_2 D_3 D_4 D_5 E_6 D_7 E_7} \bigg] \,.
\end{aligned}
\end{align}
Defining the following auxiliary superfamily
\begin{equation}
	I^{\mathrm{aux}}_{a_1 a_2 a_3 a_4 a_5 a_6 a_7 a_8 a_9} = e^{2\eps\gamma_E} \int \frac{d^d k_1}{i \pi^{d/2}} \frac{d^d k_2}{i \pi^{d/2}}
	\frac{1}{ D_1^{a_1} D_2^{a_2} D_3^{a_3} D_4^{a_4} D_5^{a_5} D_6^{a_6} D_7^{a_7} E_6^{a_8} E_7^{a_9} } \,,
\end{equation}
where $D_1,\cdots,D_7$ and $E_6,E_7$ are defined in \cref{eq:NPL1propagators} and \cref{eq:E6E7defs}, respectively,
\cref{eq:I_NPL1_scalarsubtracted} takes the following form
\begin{align}
\begin{aligned}
        \label{eq:I_NPL1_scalarsubtracted_aux}
	\tilde{G}^{\nplone}_{\mathrm{cand};\mathrm{sc}} = & \, I^{\nplone}_{111111100} - I^{\nplone}_{111111100}\big|_{\mathrm{soft}}
          - \frac{1}{2 d_{12}} \Big[ I^{\mathrm{aux}}_{1101111-11} - I^{\mathrm{aux}}_{110011101} 
	  + I^{\mathrm{aux}}_{110111001} \\ 
	& - I^{\mathrm{aux}}_{110011011} 
	  - 2(d_{12}+d_{23})\, I^{\mathrm{aux}}_{110111101} 
          + I^{\mathrm{aux}}_{01111111-1} - I^{\mathrm{aux}}_{011101110} \\
	&  + 2 d_{23}\, I^{\mathrm{aux}}_{011111110}
          + I^{\mathrm{aux}}_{011110110} - I^{\mathrm{aux}}_{011100111} \Big] \,.
\end{aligned}
\end{align}
The integrals in the auxiliary superfamily can be written in terms of integrals belonging to the master integral families 
shown in \cref{fig:MIfamilies} using IBP reduction.
We have further verified numerically using \textsc{AMFlow} that \cref{eq:I_NPL1_scalarsubtracted_aux} is indeed finite.
Finally, the following set of integrals 
\begin{equation}
	\lbrace  \tilde{G}^{\nplone}_{\mathrm{cand};\mathrm{sc}}, 
	G^{\nplone}_{\mathrm{cand};1},
	G^{\nplone}_{\mathrm{cand};2},
	G^{\nplone}_{\mathrm{cand};3}
	\rbrace \,,
\end{equation}
written explicitly in terms of the top-sector NPL1 basis through IBP relations
\begin{align}
\label{eq:finalfiniteIntegrals}
	\begin{aligned}
	\eps^4\, \tilde{G}^{\nplone}_{\mathrm{cand};\mathrm{sc}} &= \frac{1}{1+2\eps}\, \cJ^{\nplone}_{33} + \text{subsectors} \,, \\
        \eps^4\, G^{\nplone}_{\mathrm{cand};1} &= a_{34}\, \cJ^{\nplone}_{34} + \text{subsectors} \,, \\
        \eps^4\, G^{\nplone}_{\mathrm{cand};2} &= b_{34}\, \cJ^{\nplone}_{34} + b_{36}\, \cJ^{\nplone}_{36} + \text{subsectors} \,, \\
        \eps^4\, G^{\nplone}_{\mathrm{cand};3} &= \eps\, c_{33}\, \cJ^{\nplone}_{33}
                + c_{34}\, \cJ^{\nplone}_{34} + c_{35}\, \cJ^{\nplone}_{35} + c_{36}\, \cJ^{\nplone}_{36} + \text{subsectors} \,,
\end{aligned}
\end{align}
are employed to determine the elliptic master integral components $\cJ^{\nplone}_{ij}$, where $i \in \lbrace 33,34,35,36\rbrace$ and $j \in \lbrace 0,1,2,3\rbrace$,
in terms of independent polylogarithmic master integral components, which ultimately ensures analytic pole cancellation in the amplitude.
In \cref{eq:finalfiniteIntegrals}, although $\cJ^{\nplone}_{33}$ is still multiplied by $\eps$ in the last equation, it is indeed harmless since 
relations for $\cJ^{\nplone}_{33}$ components are completely fixed by the first equation.

The basis of special functions consists of the master integral components that remain in
the finite remainders after the relations of \cref{sec:polylog,sec:elliptic} have been
applied, together with those needed to close their differential equations. The procedure of
\cref{sec:polylog} gives $6$, $10$, $46$ and $106$ independent polylogarithmic functions at
weights one to four. Of these, four components of the massless one-loop master integrals of
$\qqaa$, two at weight one and two at weight three, enter only the poles,
and $29$ functions of weight four do not appear in the finite remainders. Together with the
$\eps^4$ components of the six elliptic master integrals of \cref{sec:elliptic}, this
leaves $141$ components: $22$ of one-loop master integrals, $113$ of polylogarithmic two-loop
master integrals and the six elliptic ones. Since the finite remainders also contain products
of components, we add $27$ polynomial combinations of lower-weight components as further
functions. The finite remainders are then linear in the special function monomials, and the
differential equations of the special functions form a closed linear system. Together with $1$, $\pi^2$, $\zeta_3$ and
$\pi^2$ times three weight-one functions, this gives the $174$ functions of
\cref{sec:implementation}. \Cref{tab:sfbasis} lists their number at each weight. We denote them by
$\sfn{w}{j}$, the $j$-th function of weight $w$. 
As an example, after applying the relations of \cref{sec:polylog,sec:elliptic}, the NPL1
top-sector master integral $\cJ^{\nplone}_{33}$ reads
\begin{align}
	\cJ^{\nplone}_{33} = {} & \frac{\eps^2}{2 d_{12} r_7}\, \sfn{2}{6}
	+ \frac{\eps^3}{d_{12}} \bigg[ \frac{\sfn{2}{6}}{r_7} - \frac{\sfn{1}{4}\,\sfn{2}{6}}{2 r_7}
	+ \frac{\sfn{3}{9}}{2 r_7} - \frac{4\,\sfn{3}{11}}{r_8} - \frac{4\,\sfn{3}{12}}{r_7}
	+ \frac{\sfn{3}{25}}{4 r_7} - \frac{\sfn{3}{32}}{4 r_7} \bigg] \nonumber \\
	& + \eps^4\, \sfn{4}{6} + \cO(\eps^5) \,,
	\label{eq:J33example}
\end{align}
where $m_t^2=1$ and
\begin{equation}
	r_7 = \sqrt{d_{23}(d_{12}+d_{23})\big(d_{23}^2+d_{12}d_{23}-2d_{12}\big)} \,, \qquad
	r_8 = \sqrt{-2\,d_{12}d_{23}(d_{12}+d_{23})} \,.
\end{equation}
The pole-contributing components $\cJ^{\nplone}_{33,2}$ and $\cJ^{\nplone}_{33,3}$ are expressed
through polylogarithmic special functions, while the elliptic $\eps^4$ component is kept as the
independent function $\sfn{4}{6}$.

Relations involving the $\eps^4$ components of the six elliptic master integrals have not been
searched for. Such relations may have coefficients that depend on the kinematics, so the
basis may be overcomplete in the elliptic sector.

\begin{table}[t]
	\centering
	\renewcommand{\arraystretch}{1.2}
	\begin{tabular}{lcccccc}
		\hline
		weight                                         & 0  & 1  & 2  & 3  & 4  & total\\
		\hline
		one-loop components                            & -- & 4  & 6  & 9  & 3  & 22\\
		two-loop components, polylogarithmic           & -- & -- & 4  & 35 & 74 & 113\\
		two-loop components, elliptic                  & -- & -- & -- & -- & 6  & 6\\
		product combinations                           & -- & -- & 7  & 17 & 3  & 27\\
		$1$, $\pi^2$, $\zeta_3$, $\pi^2\,\sfn{1}{j}$ & 1  & -- & 1  & 4  & -- & 6\\
		\hline
		total                                          & 1  & 4  & 18 & 65 & 86 & 174\\
		\hline
	\end{tabular}
	\caption{Number of special functions at each weight. The components are those of the master
		integrals that enter the finite remainders or are needed to close their differential
		equations, and the product combinations are polynomials in lower-weight functions. The 174
		functions satisfy the closed system of differential equations of \cref{sec:implementation}.}
	\label{tab:sfbasis}
\end{table}

\section{Implementations}
\label{sec:implementation}

After the relations among the master-integral components have been used, the finite remainders are
sums of monomials in the 174 special functions $\sfn{w}{j}$ of \cref{tab:sfbasis}, see
\cref{eq:finremSF}, with coefficients that are rational functions of $d_{12}$, $d_{23}$ and square
roots. We collect them in the vector $\sfvec$. In units of $m_t^2=1$, these functions satisfy a closed system of first-order differential
equations,
\begin{align*}
	\partial_{x}\sfvec = M_{x}(d_{12},d_{23})\,\sfvec \qquad \text{with}\quad x\in\{d_{12},d_{23}\}.
\end{align*}
The constants
$1$, $\pi^2$ and $\zeta_3$ are part of $\sfvec$, so the system is homogeneous. We obtain boundary values
for $\sfvec$ with \textsc{AMFlow}~\cite{Liu:2022chg} at 13 phase-space points of the physical region, five below and
eight above the threshold $d_{12} = 2$. They are stored with 40 digits. These values are then
transported to an arbitrary point of the physical region by integrating the system along straight
lines in the $(d_{12},d_{23})$ plane. Inside the physical region the only singular line is the top-pair
threshold $d_{12}=2$. A path that crosses it is deformed into the complex plane according to the Feynman
prescription $d_{12}+i0$. 

Given $\sfvec$ at a point, the helicity finite remainders follow from \cref{eq:finremSF}.
All square roots are evaluated on their principal branch, with the $+i0$ prescription applied when their arguments are negative.
This gives the 32 helicity finite remainders
$\tilde F^{(2)\vec h}_{\qq,t}$, $\tilde F^{(2)\vec h}_{\qq,n_t}$, $\tilde F^{(2)\vec h}_{\gg,N_c}$,
$\tilde F^{(2)\vec h}_{\gg,1/N_c}$ and $\tilde F^{(1)\vec h}_{\gg,t}$ for the independent helicity
configurations, together with their scale-dependent terms of \cref{eq:mudepQQ,eq:mudepGG}. 
The physical values of $m_t$ and $\mu$ are restored by evaluating at
$(d_{12},d_{23})/m_t^2$ with $\log\mu^2\to\log(\mu^2/m_t^2)$.
From the finite remainders we form the hard functions at the scale $\mu$, summed over helicities and
colours according to \cref{eq:hardfunctions}. 

We use two implementations to evaluate the special functions, which share the boundary values and the differential
equations, but differ in the integrator, the arithmetic and the paths. 
In both implementations, every point is computed from two different boundary points $A$ and $B$, and
we use the resulting relative error of the hard functions to estimate the number of correct digits as
\begin{equation}
	-\log_{10}\frac{|H_A-H_B|}{|H_A+H_B|} \,.
	\label{eq:estdigits}
\end{equation} 

\subsection*{\textsc{Mathematica} implementation}
The \textsc{Mathematica}-based implementation uses 
the \textsc{C++} differential equation solver of \textsc{AMFlow}, called from \textsc{Mathematica},
which expands the solution in Taylor series of order 70 at 60-digit working precision. This solver needs a
matrix that is rational in the line parameter. We achieve this with a diagonal rescaling of the basis, which removes all square roots from the differential equations. 
Points on the threshold itself are obtained from the asymptotic expansion of \textsc{AMFlow} around it. 
At a cost of about ten seconds per phase-space point, this implementation serves as our high-precision reference.
The hard functions are then obtained by evaluating the coefficients of the special function monomials in the finite remainders using rationalised values of
$(d_{12},d_{23},m_t^2,\mu^2)$ combined with the numerical values of the special functions.
A routine to evaluate the bare helicity coefficients at $m_t^2=1$ and $\mu^2 = 1$ is also provided by combining the finite remainder and the pole term.

\subsection*{\textsc{C++} implementation}
We implement the evaluation of the hard functions in a dedicated \textsc{C++} package.
In this implementation we employ a custom Bulirsch--Stoer solver written in \textsc{C++}~\cite{Bulirsch:1966,Hairer}. It
integrates with the extrapolated Gragg midpoint rule, of order 12 to 20 depending on the requested
precision, with step-size control. The DE matrix entries are exact rational combinations of only 270 functions of
$d_{12}$, $d_{23}$ and the square roots. They are evaluated by code generated for this system, in
double-double arithmetic (about 32 digits), while the solution and the extrapolation are
kept in quad-double arithmetic, so that the rounding in the integrator does not accumulate. The square roots
are evaluated numerically and their branches are tracked along the path.
The paths stay on one side of the threshold, where a boundary point is always available. Towards the
singular lines, the threshold and the edges of the physical region $d_{23}\to0$ and $d_{12}+d_{23}\to0$,
the functions grow rapidly, the transport needs many more steps and the hard functions lose accuracy.
We therefore impose a technical cut in the \textsc{C++} evaluator, such that points closer than $10^{-5}$, 
in units of $m_t^2$, to any of these lines are not evaluated, and the user should resort to the \textsc{Mathematica} evaluation routine.
The matrix evaluations are vectorised and distributed over threads for batches of phase-space points.
The expressions of finite remainders are compiled into generated code:
the 209 distinct monomials of the special functions and the square roots that appear in all helicity
coefficients are computed once per point, and the rational coefficients, in partial-fraction form, are
evaluated in double-double arithmetic. 

To balance accuracy and performance, the \textsc{C++} package employs a rescue system.
We require the computed hard functions to have at least 8 correct digits, as estimated by \cref{eq:estdigits}.
To reach the requested number of digits, the differential equations are integrated at up to three levels of internal precision, 
namely 12, 20 and 28 digits: each phase-space point is first evaluated at the lowest precision and 
promoted to the next level until all hard functions, or only the requested ones, reach the required number of correct digits.
If a point still fails at the highest level (28-digit evaluation), then the results are returned with a warning.

\subsection*{Validations}

We have performed several cross checks to validate our computation. We successfully reproduce the benchmark results for the bare helicity coefficients,
both for the $\qqaa$ and $\ggaa$ channels, provided in Ref.~\cite{Becchetti:2025rrz} at the phase-space point $(d_{12},d_{23},m_t^2,\mu^2) = (11/6,-7/12,1,1)$.
Furthermore, in the course of constructing the finite remainders the analytic pole cancellations serve as a strong check of our  computation.
The renormalisation scale dependence presented in \cref{eq:mudepQQ,eq:mudepGG} is validated by comparing to an evaluation at a rescaled phase-space point.
For a given point we rescale the kinematic inputs with $\mu^2$: $(d_{12}/\mu^2,d_{23}/\mu^2,m_t^2/\mu^2,1)$ and evaluate the finite remainder of the helicity coefficients
directly without using the scale-dependence formulae.
This calculation is done on an independent setup where the bare amplitudes are reduced numerically to a noncanonical master integral basis
automatically chosen by \textsc{LiteRed}, and the master integrals are evaluated directly using \textsc{AMFlow}.
The finite remainders are then extracted by numerically subtracting the UV and IR poles. Since the helicity coefficients are dimensionless,
the two results directly match.
We additionally compare the numerical results, for 500 randomly chosen physical points,
obtained using the \textsc{Mathematica}-based implementation against the \textsc{C++} one evaluated at the highest precision
and we find that both results agree to at least 20 estimated correct digits.

We have also compared the finite parts of the bare helicity coefficients with the series-expansion
representation of Ref.~\cite{Becchetti:2025rrz}. This representation consists of an expansion for a large top-quark mass
and of two expansions around the threshold, about $(s,t) = (4m_t^2,0)$ and $(s,u) = (4m_t^2,0)$.
The comparison was carried out at 1000 phase-space points drawn from a sample of $pp\to\gamma\gamma$ events.
We chose the points evenly in $s$, so that both the threshold region and the high-energy tail are populated.
At each point we take the expansion that agrees best with our result.
The agreement is limited only by the convergence of the expansions. Close to the threshold,
$3.8 \le s/m_t^2 \le 4.5$, all 24 helicity coefficients agree to 21--30 significant digits.
For $s/m_t^2 < 1$, where the large-mass expansion converges well, they typically agree to 14--17 digits.
In the intermediate region $1 \le s/m_t^2 \le 3$, where both types of expansion converge slowly, the agreement is 7--11 digits.
Above the threshold it decreases to 12--15 digits at $s/m_t^2 \simeq 5$, and is lost beyond $s/m_t^2 \simeq 6$--$7$,
where the truncated series no longer converge.

Both our \textsc{Mathematica} and \textsc{C++} implementations can be downloaded from the following repository:
\begin{center}
  \href{https://doi.org/10.5281/zenodo.24455284}{https://doi.org/10.5281/zenodo.24455284}
\end{center}
A description of the ancillary files and instructions for their use are provided in the \texttt{README.txt} 
file of the repository.

\section{Results}
\label{sec:results}

\begin{figure}[t]
  \centering
  \begin{subfigure}[b]{\linewidth}
    \centering
    \includegraphics[width=0.44\linewidth]{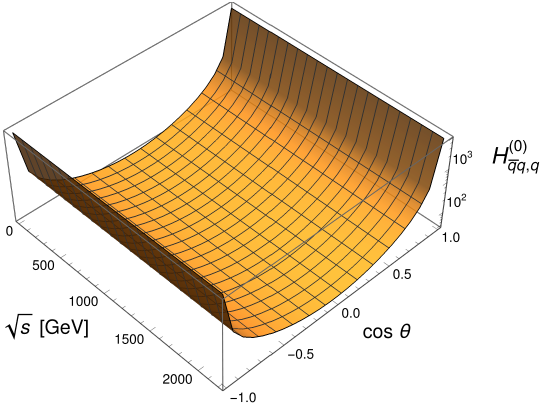} \qquad
    \includegraphics[width=0.44\linewidth]{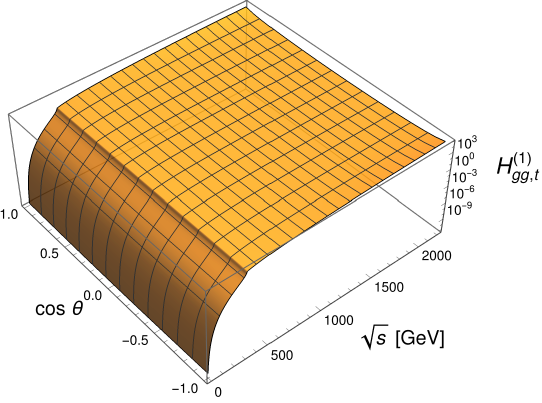}
	  \caption{The tree-level $\qqaa$ hard function $H^{(0)}_{\qq,q}$ (left) and the one-loop $\ggaa$ 
	  hard function with a closed top-quark loop, $H^{(1)}_{\gg,t}$ (right).} 
    \label{fig:H0H1plots}
  \end{subfigure}
  \\[0.7cm]
  \begin{subfigure}[b]{\linewidth}
    \centering
    \includegraphics[width=0.44\linewidth]{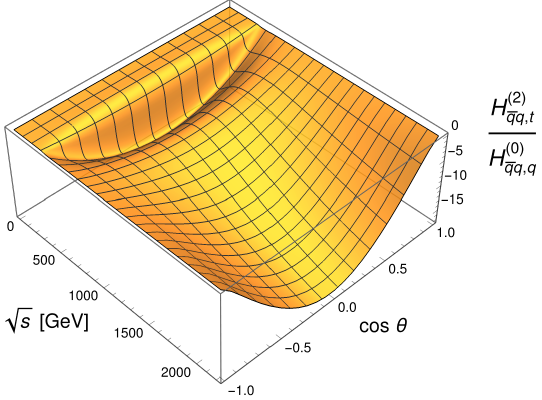} \qquad
    \includegraphics[width=0.47\linewidth]{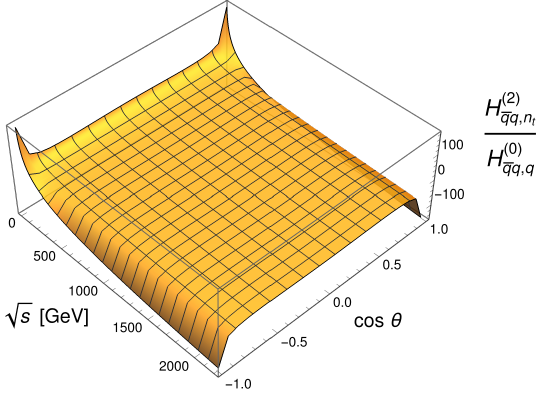} \\
    \vspace{0.7cm}
    \includegraphics[width=0.47\linewidth]{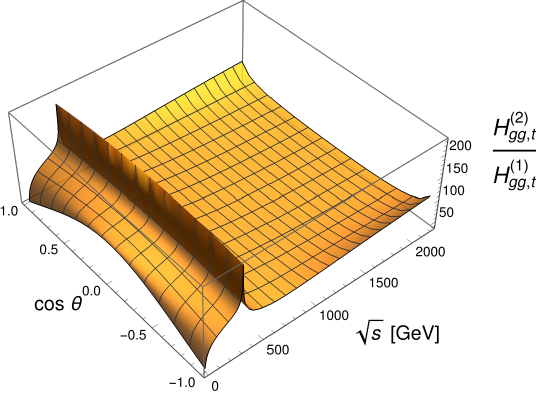}
    \caption{Two-loop hard functions for the $\qqaa$ and $\ggaa$ channels involving a top-quark loop, normalised to the corresponding lowest-order hard functions.}
    \label{fig:H2plots}
  \end{subfigure}
	\caption{Tree-level, one-loop and two-loop hard functions defined in \cref{eq:hardfunctions} evaluated on a two-dimensional grid in $(\sqrt{s},\cos\theta)$
	with $m_t = 173~\GeV$ and renormalisation scale set to the partonic centre-of-mass energy $\mu = \sqrt{s} = \sqrt{2d_{12}}$.}
  \label{fig:Hplots}
\end{figure}

In this section, we present numerical results for the $pp\to\gamma\gamma$ hard functions with a top-quark loop up to two loops, 
obtained with the \textsc{C++} implementation described in the previous section.
We first visualise in \cref{fig:Hplots} the hard functions of \cref{eq:hardfunctions},
evaluated on a two-dimensional grid of phase-space points, 
parametrised by the partonic centre-of-mass energy $\sqrt{s}$ and the scattering angle $\theta$, with
$d_{12} = \frac{s}{2}$ and $d_{23} = -\frac{s}{4}\,(1-\cos\theta)$.
Following Ref.~\cite{Becchetti:2023wev}, we consider $24$ equally spaced values of $\cos\theta$ in the range $-0.99 \le \cos\theta \le 0.99$ and 
$573$ values of $\sqrt{s}$ in the range $8~\GeV \le \sqrt{s} \le 2.2~\TeV$, for a total of $13\,752$ points. 
In contrast to Ref.~\cite{Becchetti:2023wev}, where the points are equally spaced in $s$, we take them equally spaced in $\sqrt{s}$, 
which provides a denser coverage of the low-energy and threshold regions.
The two-loop hard functions are normalised to their respective leading-order contributions and evaluated at $\mu^2 = s = 2 d_{12}$.
The hard functions $H^{(1)}_{\gg,t}$, $H^{(2)}_{\qq,t}$ and $H^{(2)}_{\gg,t}$ clearly exhibit threshold structures 
at the top-quark pair-production threshold, $\sqrt{s} = 2m_t$. In contrast, $H^{(2)}_{\qq,n_t}$ shows no such structure, 
since in this contribution the top quark enters only through a vacuum-polarisation insertion in an internal gluon propagator, 
which does not give rise to a $t\bar t$ threshold in the $s$ channel.
The enhancement of $H^{(2)}_{\qq,n_t}$ towards $\cos\theta\to\pm1$ comes from the $\log(\mu^2/m_t^2)$
terms in \cref{eq:mudepQQ} with $\mu=\sqrt{s}$, which multiply the one-loop finite remainder
$\tilde F^{(1)\vec h}_{\qq,q}$, itself enhanced by logarithms of $-d_{23}/d_{12}$ and $(d_{12}+d_{23})/d_{12}$
in the forward and backward regions. The enhancement is absent when the hard function is evaluated at $\mu=m_t$.
We note that while the $H^{(1)}_{\gg,t}$ and $H^{(2)}_{\qq,t}$ hard functions are finite at the threshold, the $H^{(2)}_{\gg,t}$ one is not,
due to divergencies in the $\tilde{F}^{(2)\vec{h}}_{\gg,N_c}$ and $\tilde{F}^{(2)\vec{h}}_{\gg,1/N_c}$ for the $++++$ and $--++$ helicity configurations.
There are two points in the two-dimensional grid that lie very close to the edges, $(\sqrt{s},\cos\theta)$ = $(8~\GeV,\pm 0.99)$, which trigger the
technical cut in the \textsc{C++} implementation. We therefore obtain the values for these two extreme points on the grid using the \textsc{Mathematica} implementation.

\begin{figure}[t]
  \centering
  \includegraphics[width=0.495\linewidth]{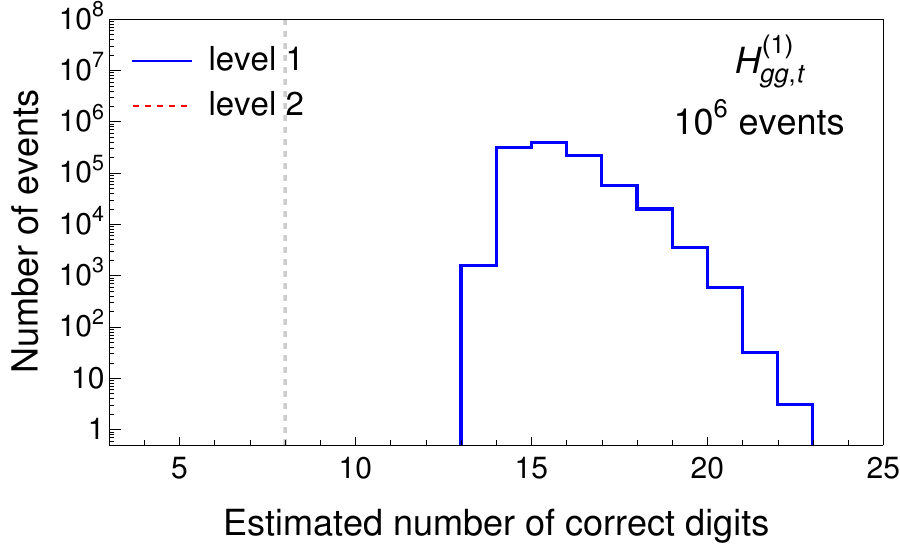} 
  \includegraphics[width=0.495\linewidth]{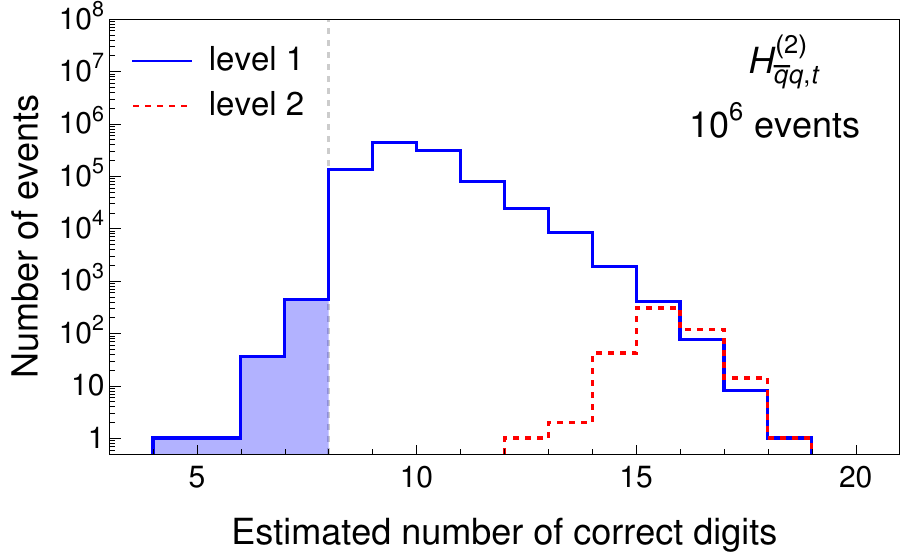} \\ 
  \includegraphics[width=0.495\linewidth]{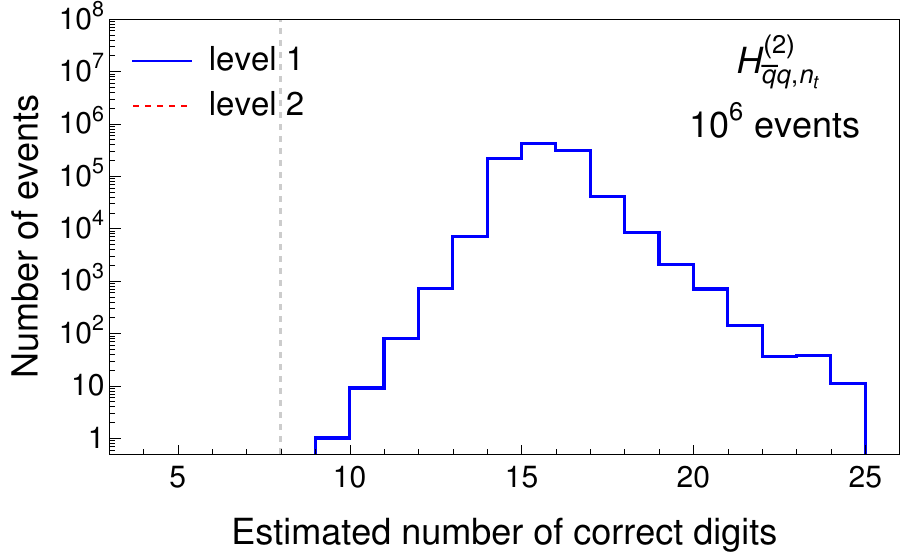} 
  \includegraphics[width=0.495\linewidth]{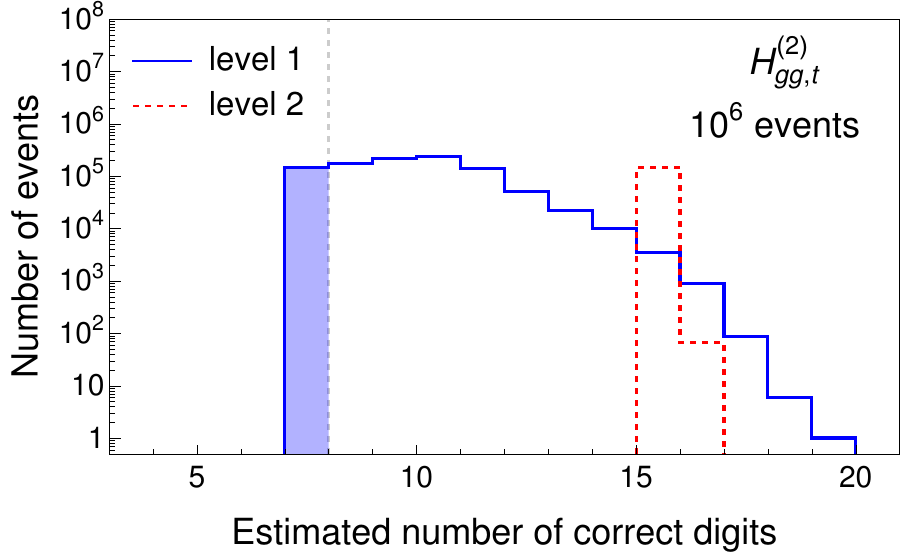} 
	\caption{Distributions of the estimated number of correct digits, \cref{eq:estdigits}, for the hard functions $H^{(1)}_{\gg,t}$, $H^{(2)}_{\qq,t}$, 
	$H^{(2)}_{\qq,n_t}$ and $H^{(2)}_{\gg,t}$, evaluated on $10^6$ phase-space points generated with \textsc{MadGraph5\_aMC@NLO}~\cite{Alwall:2014hca} with 
	standard diphoton cuts. Solid lines show the first evaluation (level 1); points below the required 8 digits (shaded, left of the dashed line) 
	are re-evaluated at the next precision level (level 2, dashed lines). We set $m_t = 173~\GeV$ and $\mu = \sqrt{s} = \sqrt{2d_{12}}$.}
  \label{fig:est_digits}
\end{figure}

To illustrate the stability and performance of the \textsc{C++} evaluator, we perform hard function evaluations on realistic collision events.
We generated one million unweighted events using \textsc{MadGraph5\_aMC@NLO}~\cite{Alwall:2014hca} for tree-level $pp\to\gamma\gamma$ production at the LHC at
a $13~\TeV$ centre-of-mass energy, where we employed the following experimental cuts on the final-state photons
\begin{equation}
	p_T(\gamma) > 40~\GeV\,, \qquad |\eta_\gamma| < 2.5 \,,\qquad \mathrm{and}\qquad \Delta R_{\gamma\gamma} > 0.4  \nonumber\,.
\end{equation}
The hard functions are evaluated with the renormalisation scale set to the partonic centre-of-mass energy, 
$\mu = \sqrt{s} = \sqrt{2d_{12}}$, and with the top-quark mass set to $m_t = 173~\GeV$.
In \cref{fig:est_digits} we present the distributions of the estimated number of correct digits, defined in \cref{eq:estdigits},
for the one- and two-loop hard functions defined in \cref{eq:hardfunctions} evaluated at one million points.
With 8 correct digits requested, we see that only the $H^{(2)}_{\qq,t}$ and $H^{(2)}_{\gg,t}$ require re-evaluation at level 2 precision,
and none of the hard functions require evaluation at level 3. For $H^{(2)}_{\qq,t}$ 0.05\% of the events require re-evaluation while for 
$H^{(2)}_{\gg,t}$ 15\% of the events require higher precision evaluation.
We note that none of the one million events triggers the technical cut.
The average evaluation time of the four hard functions (computed simultaneously for each phase-space point) is 2 ms per point on a single core,
evaluations from two different boundary points included, measured on an \textit{Intel(R) Xeon(R) Gold 6342 CPU}.

\section{Conclusions}
\label{sec:conclusions}

We have presented a computation of two-loop QCD amplitudes for diphoton production at the LHC in both the quark- and gluon-initiated channels,
focusing on the top-quark loop contributions. In contrast to the previously computed results~\cite{Becchetti:2023wev,Becchetti:2025rrz,Ahmed:2025osb},
here we derived analytic expressions for the finite remainders, valid across the full kinematic range,
which can be evaluated efficiently over the entire physical phase space. To this end,
we introduced an approach where the finite remainders of the amplitude are derived using a set of special functions built from a
noncanonical basis of elliptic master integrals.
Such a choice of basis allows us to derive the finite remainders analytically using the standard finite-field reconstruction methods,
and in addition keeps the differential equations of the special functions algebraic, allowing for a more straightforward numerical evaluation.
The special function basis is obtained by finding independent $\eps$-expanded master integral components, where we employed a numerical approach for the 
polylogarithmic sectors. The construction of the special function basis in the elliptic sectors focuses on 
relating the pole-contributing elliptic master integral components
to the polylogarithmic components, to guarantee pole cancellation in the helicity amplitudes.
The elliptic master integral components contributing only to the finite part are kept as
independent functions, so the special function basis may be overcomplete in the elliptic sector.

We constructed numerical evaluations of the hard functions by solving the differential equations
of the special functions combined with the analytic expressions of the finite remainders of the helicity amplitudes.
We provide numerical evaluation routines in \textsc{Mathematica} and \textsc{C++}.
Our \textsc{Mathematica} numerical routine also provides evaluations around and at the threshold, where finite, by implementing series expansions.
We employ the \textsc{C++} routine to evaluate one million physical events generated with realistic experimental cuts and benchmark the 
performance and stability, where, requiring an estimated precision of at least 8 significant digits on the hard functions, the average evaluation time is 2 ms per point on a single core.
Such a fast and stable implementation allows for a direct use in the cross section computation using Monte Carlo techniques.

In the two-loop massive QCD corrections to diphoton production that we considered, elliptic sectors appear only in one master integral family (NPL1).
It will be interesting to apply  the approach that is presented in this work to other scattering processes with more elliptic sectors
and higher multiplicities.

\acknowledgments
We thank Colomba Brancaccio, Dhimiter Canko, Federico Ripani and Simone Zoia for comments on the manuscript.
This work has been supported by an appointment to the JRG Program at the APCTP through the Science and Technology Promotion Fund and Lottery Fund
of the Korean Government and by the Korean Local Governments – Gyeongsangbuk-do Province and Pohang City. 
AI agents, including OpenAI Codex and Claude Code, were used under the authors’ supervision to assist
with manuscript polishing and  code implementation. The authors take full responsibility for the final manuscript
and the results it reports.

\appendix
\section{Ultraviolet and infrared poles}
\label{app:UVIRpoles}

In this Appendix we describe the construction of UV and IR counterterms to build the finite remainder in \cref{eq:finhelcoeffs}.
To remove the UV singularities, we renormalise the strong coupling in the $\overline{\rm MS}$ scheme 
and the top-quark mass and the heavy-quark contributions to the gluon and light-quark external wave functions in the on-shell scheme (OS),
\begin{align}
        \bar\alpha_s\, S_\eps &= \alpha_s\, \mu^{2\eps} \bigg[ 1 + \frac{\alpha_s}{4\pi}\, \dZ_{\alpha_s}^{(1)} + \cO(\alpha_s^2)\bigg] \,,  \\
        \bar m_t &= m_t \bigg[ 1 + \frac{\alpha_s}{4\pi}\, \dZ_{m}^{(1)} + \cO(\alpha_s^2)\bigg] ,  \\
        Z_g &= 1 + \frac{\alpha_s}{4\pi}\, \dZ_{g}^{(1)} + \cO(\alpha_s^2) \,, \\
	Z_q &= 1 + \bigg(\frac{\alpha_s}{4\pi}\bigg)^2 \dZ_{q}^{(2)} + \cO(\alpha_s^3) \,,
\end{align}
with $S_\eps = (4\pi)^\eps e^{-\eps\gamma_E}$ and 
\begin{align}
        \dZ_{\alpha_s}^{(1)} &= -\frac{\beta_0}{\eps} \,, \\
        \dZ_{m}^{(1)} &= -C_F\, e^{\eps\gamma_E}\, \Gamma(1+\eps) \left(\frac{\mu^2}{m_t^2}\right)^{\!\eps} \, \left( \frac{3}{\eps} + \frac{4}{1-2\eps} \right) \,, \\
        \dZ_{g}^{(1)} &= -\frac{4 T_F n_t}{3}\, e^{\eps\gamma_E}\, \Gamma(1+\eps) \left(\frac{\mu^2}{m_t^2}\right)^{\!\eps} \, \frac{1}{\eps} \,, \\
        \dZ_{q}^{(2)} &= C_F T_F n_t\, e^{2\eps\gamma_E}\, \Gamma^2(1+\eps) \left(\frac{\mu^2}{m_t^2}\right)^{\!2\eps}  \left( \frac{1}{\eps} - \frac{5}{6} \right) \,,
\end{align}
where
\begin{align}
	\beta_0 & = \frac{11}{3} C_A - \frac{4}{3} T_F (n_l + n_t) \,, \\
	C_F & = \frac{N_c^2-1}{2N_c} \,, \\
	C_A & = N_c \,.
\end{align}
$n_l$ and $n_t$ denote the numbers of light- and heavy-quark flavours, respectively.
The renormalised amplitudes are obtained from the bare ones as
\begin{equation}
\cA_{\rm ren} = Z_q^{n_q/2}\, Z_g^{n_g/2}\, \cA_{\rm bare}(\bar\alpha_s, \bar m_t) \,,
\end{equation}
with $n_q$ ($n_g$) the number of external massless quarks (gluons).
Up to two loops this gives, for the quark channel, whose expansion starts at tree level,
\begin{align}
        \cA^{(1)}_{\qqaa,\rm ren} &= \cA^{(1)}_{\qqaa} \,, \nonumber \\
        \cA^{(2)}_{\qqaa,\rm ren} &= \cA^{(2)}_{\qqaa} + \dZ_{\alpha_s}^{(1)}\, \cA^{(1)}_{\qqaa}
	+ \dZ_q^{(2)}\, \cA^{(0)}_{\qqaa} \,,
\end{align}
and for the loop-induced gluon channel
\begin{align}
        \cA^{(1)}_{\ggaa,\rm ren} &= \cA^{(1)}_{\ggaa} \,, \nonumber \\
        \cA^{(2)}_{\ggaa,\rm ren} &= \cA^{(2)}_{\ggaa} + \left( \dZ_{\alpha_s}^{(1)} + \dZ_g^{(1)} \right) \cA^{(1)}_{\ggaa}
	+ \dZ_m^{(1)}\, \cA^{(1),\mathrm{mct}}_{\ggaa} \,,
\end{align}
where $\cA^{(1),\mathrm{mct}}_{\ggaa}$ is the one-loop gluon-initiated amplitude with a single mass-counterterm insertion on each top-quark propagator.
The renormalisation for partial amplitudes appearing on the RHS of \cref{eq:partialamps} is obtained by filtering the relevant $N_c$, $n_t$, $Q_t$ and $Q_q$ factors.

For the IR subtraction, we follow the construction in Refs.~\cite{Catani:1998twoloopIR,Becher:2009IRPRL,Becher:2009IRstructure,
Gardi:2009softanomdim,Gardi:2009IRreview,Becher:2009kw,Ferroglia:2009ii}, where the IR poles are removed with the multiplicative factor $\ZZ$.
At two loops the finite remainders are given by
\begin{align}
        \cF^{(1)}_z &= \cA^{(1)}_{z,\rm ren} - \ZZ^{(1)}_z \cA^{(0)}_z \,, \\
	\cF^{(2)}_{z} &= \cA^{(2)}_{z,\rm ren} - \ZZ^{(1)}_z \cA^{(1)}_{z,\rm ren} - \left[ \ZZ^{(2)}_z - \big(\ZZ^{(1)}_z\big)^2 \right] \cA^{(0)}_z \,,
\end{align}
where $z$ stands for either $\qqaa$ or $\ggaa$. We will need only $\ZZ^{(1)}_{\ggaa}$ and the $n_t$ term of $\ZZ^{(2)}_{\qqaa}$  for the two-loop partial amplitudes
involving a top-quark loop considered in this paper. They are explicitly given by
\begin{align}
	\ZZ^{(1)}_{\ggaa} &= -\frac{2 C_A}{\eps^2} - \frac{1}{\eps} \left[ 2 C_A L_{12} + \hat\beta_0 \right] \,, \\
	\ZZ^{(2)}_{\qqaa}\big|_{n_t} & = -\frac{4 C_F T_F}{3} \bigg[ \frac{2 L_m}{\eps^2} + \frac{(2L_{12} + 3) L_m - L_m^2 - \zeta_2}{\eps}
        \nonumber \\
	& \quad + \frac{L_m^3}{3} + \zeta_2 L_m + \frac{2\zeta_3}{3} - \frac{1}{2}\left(2L_{12} + 3\right)\left(L_m^2 + \zeta_2\right) \bigg] + \cO(\eps) \,,
\end{align}
where 
\begin{align}
	L_m &= \log\left(\frac{m_t^2}{\mu^2}\right) \,, \\
	L_{12} &= \log\left(\frac{\mu^2}{-2d_{12}}\right) \,, \\
	\hat\beta_0 &= \frac{11}{3} C_A - \frac{4}{3} T_F n_l  \,.
\end{align}

\section{Relations among master integral components: a one-loop example}
\label{app:oneloop}

In this Appendix we use the ten one-loop master integrals with a top-quark loop, which
enter the finite remainders of $\ggaa$ together with the two-loop ones, to illustrate the procedure to find the relations among the components of the master
integrals as discussed in \cref{sec:polylog}. The system is small, but it contains
the main features of the procedure, which applies in the same way to larger systems.

The integrals of a family $F$ are
\begin{equation}
  J_F(a_1,a_2,a_3,a_4)=e^{\eps\gamma_E}\int\frac{d^dk}{i\pi^{d/2}}\,
  \frac{1}{D_1^{a_1}D_2^{a_2}D_3^{a_3}D_4^{a_4}},\qquad d=4-2\eps ,
  \label{eq:1Lfamily}
\end{equation}
with the inverse propagators $D_i$ of \cref{tab:1Lfamilies}. The subscript
of a family gives the permutations of the external legs.
\begin{table}[t]
\centering
\renewcommand{\arraystretch}{1.2}
\begin{tabular}{lcccc}
\hline
family & $D_1$ & $D_2$ & $D_3$ & $D_4$\\
\hline
$J_{1234}$ & $k^2-m_t^2$ & $(k-p_1)^2-m_t^2$ & $(k-p_1-p_2)^2-m_t^2$ & $(k+p_4)^2-m_t^2$\\
$J_{1243}$ & $k^2-m_t^2$ & $(k-p_1)^2-m_t^2$ & $(k-p_1-p_2)^2-m_t^2$ & $(k+p_3)^2-m_t^2$\\
$J_{1324}$ & $k^2-m_t^2$ & $(k-p_1)^2-m_t^2$ & $(k-p_1-p_3)^2-m_t^2$ & $(k+p_4)^2-m_t^2$\\
\hline
\end{tabular}
\caption{Inverse propagators of the one-loop integral families.}
\label{tab:1Lfamilies}
\end{table}

The ten master integrals are a tadpole $\Tad$, three bubbles $\Bub_x$, three triangles
$\Tri_x$ and three boxes $\Boxm_{xy}$. They read
\begin{equation}
\begin{aligned}
  &\Tad=\eps\,J_{\Tad},\qquad J_{\Tad}\equiv J_{1324}(0,0,0,2),\\
  &\Bub_s=\eps\,r_s\,J_{1243}(1,0,2,0),\\
  &\Bub_t=\eps\,r_t\,J_{1324}(0,1,0,2),\\
  &\Bub_u=\eps\,r_u\,J_{1324}(1,0,2,0),\\
  &\Tri_s=-\eps^2J_{\Tad}+\eps\,(d_{12}-2m_t^2)\,J_{1243}(1,0,2,0)
          +\eps\,m_t^2\,d_{12}\,J_{1243}(1,0,1,2),\\
  &\Tri_t=-\eps^2J_{\Tad}+\eps\,(d_{23}-2m_t^2)\,J_{1324}(0,1,0,2)
          +\eps\,m_t^2\,d_{23}\,J_{1324}(0,1,2,1),\\
  &\Tri_u=\phantom{-}\eps^2J_{\Tad}-\eps\,(d_{13}-2m_t^2)\,J_{1324}(1,0,2,0)
          -\eps\,m_t^2\,d_{13}\,J_{1324}(1,0,1,2),\\
  &\Boxm_{st}=\eps^2\,r_{st}\,J_{1234}(1,1,1,1),\\
  &\Boxm_{su}=\eps^2\,r_{su}\,J_{1243}(1,1,1,1),\\
  &\Boxm_{tu}=\eps^2\,r_{tu}\,J_{1324}(1,1,1,1).
\end{aligned}
\label{eq:1Lmasters}
\end{equation}
The relevant square roots are
\begin{equation}
\begin{gathered}
  r_s=\sqrt{d_{12}(d_{12}-2m_t^2)},\qquad
  r_t=\sqrt{d_{23}(d_{23}-2m_t^2)},\qquad
  r_u=\sqrt{d_{13}(d_{13}-2m_t^2)},\\
  r_{st}=\sqrt{d_{12}d_{23}\big[d_{12}d_{23}-2m_t^2(d_{12}+d_{23})\big]},\qquad
  r_{su}=\sqrt{d_{12}d_{13}\big[d_{12}d_{13}-2m_t^2(d_{12}+d_{13})\big]},\\
  r_{tu}=\sqrt{d_{13}d_{23}\big[d_{13}d_{23}-2m_t^2(d_{13}+d_{23})\big]},
\end{gathered}
\label{eq:1Lroots}
\end{equation}
with $d_{13}=-d_{12}-d_{23}$.

From here on we set $m_t^2=1$. The ten master integrals satisfy canonical differential
equations of the form of \cref{eq:canonicalform}, and their components satisfy
\cref{eq:compDE}. The components are not independent, and we now determine the
relations among them with the procedure of \cref{sec:polylog}. Its only inputs are
\cref{eq:compDE} and the numerical values of the components at a reference phase-space point.

We write $X^{(w)}$ for the weight-$w$ component of a master integral $X$, i.e.\ $\cJ_{iw}$ in
\cref{eq:MIexpansion}. The weight-zero components are rational numbers, which are read off
from the numerical values. The only non-zero one is $\Tad^{(0)}=1$. At each weight $w=1,\dots,4$, the
components whose derivative vanishes identically are constants and are fixed from the
numerical values. The other components and the products of lower-weight basis functions
form the ansatz of \cref{eq:ansatz}, i.e.\ the columns of $\mathbf N$, which we order as
\begin{equation}
	\big(\,\Boxm_{tu},\Boxm_{su},\Boxm_{st}\;\big|\;\Tri_u,\Tri_t,\Tri_s\;\big|\;
	\Bub_u,\Bub_t,\Bub_s\;\big|\;\Tad\;\big|\;\text{products}\,\big),
	\label{eq:1Lorder}
\end{equation}
so that boxes and triangles are expressed through bubbles, and the products, which are
linearly independent among themselves, are never eliminated. The matrix $\mathbf N$ is built at
$n_P=6$ points $P_1,\dots,P_6$ close to the reference point.\\

\noindent\underline{\textit{Weight one}}\\
Since only $\Tad^{(0)}$ is non-zero, only the bubbles, whose equations
contain $\Tad$, have non-vanishing derivatives. Writing
$\nabla=(\partial_{d_{12}},\partial_{d_{23}})$,
\begin{equation}
  \nabla\Bub_s^{(1)}=\Big(\frac1{2r_s},\,0\Big),\qquad
  \nabla\Bub_t^{(1)}=\Big(0,\,\frac1{2r_t}\Big),\qquad
  \nabla\Bub_u^{(1)}=-\frac1{2r_u}\,(1,1).
  \label{eq:1Lw1}
\end{equation}
The derivatives of $\Tad^{(1)}$, $\Tri_x^{(1)}$ and $\Boxm_{xy}^{(1)}$ vanish. These seven
components are constants. Their values at the reference point vanish, so
$\Tad^{(1)}=\Tri_x^{(1)}=\Boxm_{xy}^{(1)}=0$.

The remaining three components are the columns of $\mathbf N$. Their derivatives contain no
lower-weight function, so the only row label is $g=1$, and each point contributes two rows,
the values of $\partial_{d_{12}}$ and $\partial_{d_{23}}$. With the columns in the order of
\cref{eq:1Lorder}, $\big(\Bub_u^{(1)},\Bub_t^{(1)},\Bub_s^{(1)}\big)$,
\cref{eq:1Lw1} gives
\begin{equation}
  \mathbf N=\begin{pmatrix}\mathbf N(P_1)\\ \vdots\\ \mathbf N(P_6)\end{pmatrix},\qquad
  \mathbf N(P)=\begin{pmatrix}
    -\dfrac{1}{2r_u} & 0 & \dfrac{1}{2r_s}\\[2ex]
    -\dfrac{1}{2r_u} & \dfrac{1}{2r_t} & 0
  \end{pmatrix}_{\!P},
  \label{eq:1LMw1}
\end{equation}
where the first row of $\mathbf N(P)$ is $\partial_{d_{12}}$, the second $\partial_{d_{23}}$, and the
roots are evaluated at the point $P$. The three gradients in \cref{eq:1Lw1} are
linearly independent, so $\mathbf N$, of size $12\times3$, has full rank and no null vector.
The weight-one basis is
\begin{equation}
  S_1=\big\{\Bub_s^{(1)},\,\Bub_t^{(1)},\,\Bub_u^{(1)}\big\}.
\end{equation}\\

\noindent\underline{\textit{Weight two}}\\
The derivative of $\Tad^{(2)}$ vanishes, and the numerical value gives
$\Tad^{(2)}=\pi^2/12$. The ansatz consists of the other $9$ components and the $6$ products
of two functions of $S_1$, a total of $15$ elements. Every derivative is a linear combination of the three functions
of $S_1$, which are the row labels, $n_g=3$. Each point thus contributes $6$ rows, and $\mathbf N$
has size $36\times15$.

Since $\Tad^{(1)}$, $\Tri_x^{(1)}$ and $\Boxm_{xy}^{(1)}$ vanish, only the bubble columns of
$A_x$ contribute to the derivatives of the weight-two components, and the derivatives of the
products follow from \cref{eq:1Lw1}. We abbreviate $b_x\equiv\Bub_x^{(1)}$ and order the
six rows of a point as $(\partial_{d_{12}};\,b_u,b_t,b_s)$ and
$(\partial_{d_{23}};\,b_u,b_t,b_s)$. The block $\mathbf N(P)$ of a point $P$, which contains the
values $R^{(x)}_{kj}(P)$, then consists of the nine columns of the components,
$\mathbf N_{\rm comp}(P)$,
\begin{equation}
\renewcommand{\arraystretch}{1.8}\setlength{\arraycolsep}{3.5pt}
\begin{array}{cc|ccc|ccc|ccc}
 & & \Boxm_{tu}^{(2)} & \Boxm_{su}^{(2)} & \Boxm_{st}^{(2)}
   & \Tri_u^{(2)} & \Tri_t^{(2)} & \Tri_s^{(2)}
   & \Bub_u^{(2)} & \Bub_t^{(2)} & \Bub_s^{(2)}\\ \hline
 & b_u & -\frac{2d_{23}}{r_u\,r_{tu}} & \frac{d_{13}^2+2d_{23}}{r_u\,r_{su}} & 0
   & -\frac{1}{r_u} & 0 & 0 & -\frac{1}{2-d_{13}} & 0 & 0\\
 \partial_{d_{12}} & b_t & -\frac{r_t}{r_{tu}} & 0 & \frac{r_t}{r_{st}}
   & 0 & 0 & 0 & 0 & 0 & 0\\
 & b_s & 0 & -\frac{d_{12}^2+2d_{23}}{r_s\,r_{su}} & \frac{2d_{23}}{r_s\,r_{st}}
   & 0 & 0 & -\frac{1}{r_s} & 0 & 0 & \frac{1}{2-d_{12}}\\ \hline
 & b_u & \frac{d_{13}^2+2d_{12}}{r_u\,r_{tu}} & -\frac{2d_{12}}{r_u\,r_{su}} & 0
   & -\frac{1}{r_u} & 0 & 0 & -\frac{1}{2-d_{13}} & 0 & 0\\
 \partial_{d_{23}} & b_t & -\frac{d_{23}^2+2d_{12}}{r_t\,r_{tu}} & 0 & \frac{2d_{12}}{r_t\,r_{st}}
   & 0 & -\frac{1}{r_t} & 0 & 0 & \frac{1}{2-d_{23}} & 0\\
 & b_s & 0 & -\frac{r_s}{r_{su}} & \frac{r_s}{r_{st}}
   & 0 & 0 & 0 & 0 & 0 & 0
\end{array}
\label{eq:1LMw2comp}
\end{equation}
and the six columns of the products, $\mathbf N_{\rm prod}(P)$,
\begin{equation}
\renewcommand{\arraystretch}{1.6}\setlength{\arraycolsep}{8pt}
\begin{array}{cc|cccccc}
 & & b_u^2 & b_t^2 & b_s^2 & b_t\,b_u & b_s\,b_u & b_s\,b_t\\ \hline
 & b_u & -\frac{1}{r_u} & 0 & 0 & 0 & \frac{1}{2r_s} & 0\\
 \partial_{d_{12}} & b_t & 0 & 0 & 0 & -\frac{1}{2r_u} & 0 & \frac{1}{2r_s}\\
 & b_s & 0 & 0 & \frac{1}{r_s} & 0 & -\frac{1}{2r_u} & 0\\ \hline
 & b_u & -\frac{1}{r_u} & 0 & 0 & \frac{1}{2r_t} & 0 & 0\\
 \partial_{d_{23}} & b_t & 0 & \frac{1}{r_t} & 0 & -\frac{1}{2r_u} & 0 & 0\\
 & b_s & 0 & 0 & 0 & 0 & -\frac{1}{2r_u} & \frac{1}{2r_t}
\end{array}
\label{eq:1LMw2prod}
\end{equation}

Thus
\begin{equation}
  \mathbf N(P)=\big(\,\mathbf N_{\rm comp}(P)\;\big|\;\mathbf N_{\rm prod}(P)\,\big),
\end{equation}
and, as at weight one, $\mathbf N$ is the stack of the six blocks $\mathbf N(P_n)$.

The matrix $\mathbf N$ has rank $12$, and the relations follow from the reduced row echelon form of
its null space. They read
\begin{equation}
  \Tri_s^{(2)}=-\big(\Bub_s^{(1)}\big)^2 +c_1,\qquad
  \Tri_t^{(2)}=-\big(\Bub_t^{(1)}\big)^2 +c_2,\qquad
  \Tri_u^{(2)}=\big(\Bub_u^{(1)}\big)^2 +c_3,
  \label{eq:1Lw2rel}
\end{equation}
where the numerical values give $c_1=c_2=c_3=0$.
No null vector involves the boxes or the bubbles, so the weight-two basis is
\begin{equation}
  S_2=\big\{\Boxm_{st}^{(2)},\,\Boxm_{su}^{(2)},\,\Boxm_{tu}^{(2)},\,
  \Bub_s^{(2)},\,\Bub_t^{(2)},\,\Bub_u^{(2)}\big\}.
\end{equation}\\

\noindent\underline{\textit{Weight three}}\\
The constant is $\Tad^{(3)}=-\zeta_3/3$. The ansatz consists of $9$ components and $31$
products: $10$ cubic monomials in $S_1$, $18$ products of a function of $S_1$ with one of
$S_2$, and $\pi^2$ times the three functions of $S_1$. The row labels are the $13$ monomials
of weight two: the six functions of $S_2$, the six quadratic monomials in $S_1$, and
$\pi^2$. The derivatives
of the boxes $\Boxm_{xy}^{(3)}$ contain the triangle components $\Tri_x^{(2)}$ and
$\Tri_y^{(2)}$,  which are not basis functions and are replaced by
\cref{eq:1Lw2rel}. The derivatives of the bubbles contain the tadpole $\Tad^{(2)}=\pi^2/12$, which is the
origin of the row label $\pi^2$. This is why the constants of weight $w$ have to be fixed before weight $w+1$ is treated.
The matrix has size $156\times40$ and full rank. Thus, the weight-three basis consists of all nine non-constant components,
\begin{equation}
  S_3=\big\{\Boxm_{xy}^{(3)},\,\Tri_x^{(3)},\,\Bub_x^{(3)}\big\}.
\end{equation}\\

\noindent\underline{\textit{Weight four}}\\
The constant is $\Tad^{(4)}=\pi^4/160$. The ansatz consists of $9$ components and $114$
products, there are $n_g=41$ row labels, and $\mathbf N$ has size $492\times123$. The rank of $\mathbf N$ is
$120$. The three null vectors give the triangles,
\begin{equation}
\begin{aligned}
  \Tri_s^{(4)}&=\big(\Bub_s^{(2)}\big)^2-2\,\Bub_s^{(1)}\Bub_s^{(3)}
     +\frac{\pi^2}{12}\big(\Bub_s^{(1)}\big)^2 + c_1,\\
  \Tri_t^{(4)}&=\big(\Bub_t^{(2)}\big)^2-2\,\Bub_t^{(1)}\Bub_t^{(3)}
     +\frac{\pi^2}{12}\big(\Bub_t^{(1)}\big)^2+ c_2,\\
  \Tri_u^{(4)}&=-\big(\Bub_u^{(2)}\big)^2+2\,\Bub_u^{(1)}\Bub_u^{(3)}
     -\frac{\pi^2}{12}\big(\Bub_u^{(1)}\big)^2 + c_3.
\end{aligned}
\end{equation}
These relations connect components of three different weights. The numerical values give
again $c_1=c_2=c_3=0$. The weight-four basis is
\begin{equation}
  S_4=\big\{\Boxm_{st}^{(4)},\,\Boxm_{su}^{(4)},\,\Boxm_{tu}^{(4)},\,
  \Bub_s^{(4)},\,\Bub_t^{(4)},\,\Bub_u^{(4)}\big\}.
\end{equation}

\Cref{tab:1Lexample} collects the numbers of the four steps. \Cref{tab:1Lmasters} lists the
independent components and their names as special functions.
\begin{table}[t]
\centering
\renewcommand{\arraystretch}{1.2}
\begin{tabular}{lcccc}
\hline
weight $w$                                   & 1     & 2      & 3       & 4\\
\hline
components                                   & 10    & 10     & 10      & 10\\
constants                                    & 7     & 1      & 1       & 1\\
columns of $\mathbf N$ (components $+$ products)     & $3+0$ & $9+6$  & $9+31$  & $9+114$\\
row labels $n_g$                             & 1     & 3      & 13      & 41\\
size of $\mathbf N$ ($n_P=6$)                          & $12\times3$ & $36\times15$ & $156\times40$ & $492\times123$\\
rank $r$                                     & 3     & 12     & 40      & 120\\
relations (null space dimension)             & 0     & 3      & 0       & 3\\
independent functions                        & 3     & 6      & 9       & 6\\
\hline
\end{tabular}
\caption{The relation search for the ten massive one-loop master integrals.}
\label{tab:1Lexample}
\end{table}
\begin{table}[t]
\centering
\renewcommand{\arraystretch}{1.2}
\begin{tabular}{lr|cccc}
\hline
 & $w=0$ & $w=1$ & $w=2$ & $w=3$ & $w=4$\\
\hline
$\Tad$         & $1$  & -- & -- & -- & --\\
$\Bub_s$       & $0$  & $\sfn{1}{1}$ & $\sfn{2}{2}$ & $\sfn{3}{2}$ & $*$\\
$\Bub_t$       & $0$  & $\sfn{1}{3}$ & $\sfn{2}{5}$ & $\sfn{3}{6}$ & $*$\\
$\Bub_u$       & $0$  & $\sfn{1}{2}$ & $\sfn{2}{4}$ & $\sfn{3}{5}$ & $*$\\
$\Tri_s$       & $0$  & -- & -- & $\sfn{3}{3}$ & --\\
$\Tri_t$       & $0$  & -- & -- & $\sfn{3}{8}$ & --\\
$\Tri_u$       & $0$  & -- & -- & $\sfn{3}{7}$ & --\\
$\Boxm_{st}$   & $0$  & -- & $\sfn{2}{1}$ & $\sfn{3}{1}$ & $\sfn{4}{1}$\\
$\Boxm_{su}$   & $0$  & -- & $\sfn{2}{3}$ & $\sfn{3}{4}$ & $\sfn{4}{2}$\\
$\Boxm_{tu}$   & $0$  & -- & $\sfn{2}{6}$ & $\sfn{3}{9}$ & $\sfn{4}{3}$\\
\hline
\end{tabular}
\caption{The massive one-loop master integrals, their weight-zero components, and the names
of their components of weights one to four that are special functions. A dash marks a
component that is a constant or the left-hand side of a relation, and an asterisk an
independent component that does not appear in the finite remainders.}
\label{tab:1Lmasters}
\end{table}

\newpage

\bibliographystyle{JHEP}
\bibliography{diphoton_2l_toploop}
\end{document}